\documentclass[aps,prd,onecolumn,eqsecnum,amsmath,nofootinbib,preprintnumbers]{revtex4}%

\usepackage{color,graphicx,float,subfigure,xcolor}
\usepackage{amsfonts,amssymb,theorem,mathrsfs,times}
\usepackage{bm}
\usepackage{multirow}
\usepackage{mathtools}
\usepackage{amsfonts,amssymb,theorem,mathrsfs}
\usepackage{dsfont}
\usepackage{setspace}
\usepackage{amsmath} 
\usepackage[colorlinks,linkcolor=blue,anchorcolor=blue,citecolor=blue]{hyperref}   
\usepackage{ulem}   
\usepackage[english]{babel}
\usepackage{physics}

\newcommand{\Arg}{\operatorname{Arg}}

{\theorembodyfont{\upshape}
	}
{\theorembodyfont{\upshape}
	}
{\theorembodyfont{\upshape}
	}
{\theorembodyfont{\upshape}
	}
{\theorembodyfont{\upshape}
	}
{\theorembodyfont{\upshape}
	}
\addtocounter{MaxMatrixCols}{10}
\newcommand{\dalm}{\kern1pt\vbox{\hrule height 0.9pt\hbox{\vrule width 0.9pt\hskip 2.5pt\vbox{\vskip 5.5pt}\hskip 3pt\vrule width 0.3pt}\hrule height 0.3pt}\kern1pt}

\begin{document}
\thispagestyle{empty}
\title{Waveform stability for the piecewise step approximation of Regge-Wheeler potential}
	
%


\author{Liang-Bi Wu$^{a\, ,c}$\footnote{e-mail address: liangbi@mail.ustc.edu.cn}}

\author{Libo Xie$^{a\, ,b\, ,c}$\footnote{e-mail address: xielibo23@mails.ucas.ac.cn (corresponding author)}}

 
\author{Yu-Sen Zhou$^d$\footnote{e-mail address: zhou\_ys@mail.ustc.edu.cn}}	

\author{Zong-Kuan Guo$^{b\, ,a\, ,c}$\footnote{e-mail address: guozk@itp.ac.cn}}

\author{Rong-Gen Cai$^{e\, ,a}$\footnote{e-mail address: cairg@itp.ac.cn}}

\affiliation{${}^a$School of Fundamental Physics and Mathematical Sciences, Hangzhou Institute for Advanced Study, UCAS, Hangzhou 310024, China}

\affiliation{${}^b$Institute of Theoretical Physics, Chinese Academy of Sciences, Beijing 100190, China}

\affiliation{${}^c$University of Chinese Academy of Sciences, Beijing 100049, China}

\affiliation{${}^d$Interdisciplinary Center for Theoretical Study and Department of Modern Physics, University of Science and Technology of China, Hefei, Anhui 230026, China}

\affiliation{${}^e$Institute of Fundamental Physics and Quantum Technology, Ningbo University, Ningbo 315211, China}

\date{\today}
	
\begin{abstract}
By interpreting the difference between the original Regge-Wheeler potential and its piecewise step approximation as perturbative effects induced by the external environment of the black hole, we investigate the stability of Schwarzschild black hole time-domain waveforms. In this work, we employ the Green's function method under the assumption of an observer at spatial infinity to obtain the waveform. For two cases in which the initial Gauss bump near the event horizon or at spatial infinity, we derive analytic expressions for the corresponding waveforms. Our results demonstrate that the waveform is indeed insensitive to tiny modifications of the effective potential, thereby confirming its stability. More importantly, we find that broader initial bumps imprint the influence of small environmental modifications more clearly on the waveform, which may provide theoretical guidance for probing the exterior environment of black holes.
\end{abstract}

\maketitle
	
\section{Introduction}\label{Introduction}
Upon characterizing the intrinsic oscillatory behavior of perturbed black holes, quasinormal modes (QNMs) manifest themselves as a discrete set of complex frequencies. The oscillation frequency is encoded in the real part, and the decay rate in the imaginary part. In the aftermath of compact object mergers, like binary black hole coalescence, the remnant undergoes ringdown. This relaxation process produces gravitational radiation overwhelmingly dominated by QNMs, whose damped oscillations are determined purely by the final spacetime geometry. Thus, QNMs not only act as spectral fingerprints that identify black holes, but also offer a powerful probe to test the Kerr hypothesis and explore gravity in the strong-field regime. This program is known as black hole spectroscopy~\cite{Kokkotas:1999bd,Berti:2009kk,Konoplya:2011qq,Berti:2025hly}.

The QNM spectra of black holes exhibit the instability~\cite{Konoplya:2022pbc,Jaramillo:2020tuu,Shen:2025yiy} due to the non-Hermiticity of black hole systems~\cite{Ashida:2020dkc}. In other words, these spectra will shift disproportionately far in the complex plane in response to seemingly minor external perturbations. Initial studies of QNM spectrum instability were provided by Nollert and Price~\cite{Nollert:1996rf,Nollert:1998ys}. Two primary methodological categories are utilized for examining spectrum instability associated with QNMs. One of the methods involves modifying the effective potential~\cite{Qian:2020cnz,Daghigh:2020jyk,Liu:2021aqh,Li:2024npg,Qian:2024iaq,Xie:2025jbr,Berti:2022xfj,Cheung:2021bol,Yang:2024vor,Courty:2023rxk,Cardoso:2024mrw,Ianniccari:2024ysv,MalatoCorrea:2025iuc,Kyutoku:2022gbr,Mai:2025cva,Yang:2025dbn}, a method whose applications are not confined to the study of instability. Furthermore, based on the smoothness of the modified effective potential, the above studies can be further classified. On the one hand, Refs. \cite{Qian:2020cnz,Daghigh:2020jyk,Liu:2021aqh,Li:2024npg,Qian:2024iaq,Xie:2025jbr} focus on non-smooth modifications of the effective potential to study the instability of QNM spectra. Such non-smooth modifications of the effective potential can have a profound impact on the QNMs, as already emphasized in~\cite{Ching:1993gt}. On the other hand, Refs. \cite{Berti:2022xfj,Cheung:2021bol,Yang:2024vor,Courty:2023rxk,Cardoso:2024mrw,Ianniccari:2024ysv,MalatoCorrea:2025iuc,Kyutoku:2022gbr} investigate the instability of QNM spectra by adding small smooth bump to the original effective potential. When the small extra bump becomes negative enough, spacetime will even become unstable~\cite{Mai:2025cva}. Notably, adding a small bump perturbation to the ordinary effective potential enables the discovery of exceptional points (EPs), in which two QNM spectra merge into one. In the vicinity of such EPs, the conventional superposition of QNMs may become invalid~\cite{Yang:2025dbn}. Alternatively, one can also change the boundary conditions of QNMs to study spectrum instability~\cite{Solidoro:2024yxi,Oshita:2025ibu,Destounis:2025dck}. These modified boundary conditions provide an effective description of spacetime deformations that modify the effective potential at remote distance. Consequently, this prescription yields results equivalent to those achieved by the explicit inclusion of a bump-like perturbation within the master equation~\cite{Oshita:2025ibu}. Pseudospectrum analysis~\cite{trefethen2020spectra}, as the second method, is also utilized to investigate the instability of the QNM spectrum~\cite{Boyanov:2024fgc, Jaramillo:2020tuu,Destounis:2023ruj,Jaramillo:2021tmt}. This approach focuses on examining the characteristic properties of non-self-adjoint operators in dissipative systems, employing visual methods to elucidate their spectrum instability. In the context of gravity theory, pseudospectra have been employed as qualitative indicators of spectrum instability in diverse spacetimes~\cite{Jaramillo:2020tuu,Destounis:2021lum,Cao:2024oud,Cao:2024sot,Arean:2024afl,Garcia-Farina:2024pdd,Arean:2023ejh,Boyanov:2023qqf,Cownden:2023dam,Carballo:2025ajx,Sarkar:2023rhp,Destounis:2023nmb,Luo:2024dxl,Warnick:2024usx,Chen:2024mon,Boyanov:2022ark,Siqueira:2025lww,Besson:2024adi,dePaula:2025fqt,Cai:2025irl,Cao:2025qws}. Transient dynamics related to pseudospectra are studied in~\cite{Carballo:2024kbk,Jaramillo:2022kuv,Chen:2024mon,Carballo:2025ajx,Besson:2025ghu}.

While frequency-domain analyses reveal that small disturbances can cause substantial spectral changes, as demonstrated by studies of effective potential modifications and pseudospectrum approaches, the ringdown waveform remains stable and is affected only perturbatively by potential modifications~\cite{Berti:2022xfj,Daghigh:2020jyk,Yang:2024vor,Oshita:2025ibu}. Mathematically, spectrum instability arises from the instability of the complex zeros of the analytically continued function $A_{\text{in}}(\omega)$, whereas waveform stability is ensured by the stability of its values on the real axis, since the Green’s function is proportional to $1/A_{\text{in}}(\omega)$ for real $\omega$~\cite{Kyutoku:2022gbr}. As a result, it was proposed that the black hole graybody factors are more robust observables~\cite{Oshita:2023cjz,Rosato:2024arw,Oshita:2024fzf,Ianniccari:2024ysv,Wu:2024ldo,Konoplya:2025ixm,Xie:2025jbr,Kyutoku:2022gbr} than the QNM spectra. It is recognized that the environmental effects surrounding a black hole induce not merely alterations to a single localized bump, but rather modifications throughout various positions of the effective potential. Furthermore, considering both simplicity and practicality, piecewise step~\cite{Nollert:1996rf}, linear~\cite{Daghigh:2020jyk}, and quadratic~\cite{Xie:2025jbr} approximations of the effective potential constitute feasible and reliable approaches to mimic modifications at numerous positions within the potential. Here, in this work, we study the stabilities of reflection amplitude $A_{\text{out}}(\omega)/A_{\text{in}}(\omega)$ and transmission amplitude $1/A_{\text{in}}(\omega)$ for the piecewise step approximation of Regge-Wheeler (R-W) potential. More explicitly, it is not only the modules of these two quantities, but also the phases of these two quantities are considered~\cite{Kyutoku:2022gbr}.

In general, it is known that the ringdown waveform can be expressed as a superposition of QNMs by the residue theorem, it takes the form $\sum_{n} E_{n}T_{n}\mathrm{e}^{-\mathrm{i}\omega_{n} t}$, where the coefficient $E_n$ is called the excitation factors (EFs)~\cite{Leaver:1986gd,Sun:1988tz,Andersson:1995zk,Glampedakis:2003dn,Berti:2006wq,Zhang:2013ksa,Silva:2024ffz,Oshita:2024wgt,Oshita:2025ibu,Lo:2025njp,Takahashi:2025uwo,Oshita:2021iyn,Chen:2024hum,Motohashi:2024fwt,Kubota:2025hjk} and $T_n$ is called the source factors. Distinct sources yield distinct waveforms, thereby influencing their stabilities. Particularly, for delta function source, waveform stability was investigated in~\cite{Oshita:2025ibu} through an approach involving modification on the QNMs boundary conditions. Leveraging the functions $A_{\text{in}}(\omega)$ and $A_{\text{out}}(\omega)$ derived from the piecewise step effective potential, we thereby investigate waveform stability not only under delta function sources but also for Gauss bump sources, where the sources are placed near the event horizon or at spatial infinity for simplicity. The time-domain stability is characterized by the mismatch function. Furthermore, we investigate the impact of the Gauss bump source's width and found that a wider initial Gauss bump is more effective at capturing minor modifications on the effective potential.

The paper is organized as follows. In Sec. \ref{Piecewise_step_approximation}, we demonstrate how the R-W potential can be approximated by a piecewise step potential. In Sec. \ref{sec: QNMs}, we use the transfer matrix formalism to solve the QNM spectra for the piecewise step potential. In Sec. \ref{sec: EFs}, we briefly revisit the relevant Green's function method and get the EFs for the piecewise step potential. In Sec. \ref{sec: stability_of_waveforms}, we study the stability of the waveform constructed from both Dirac delta source and Gauss bump source. Sec. \ref{sec: conclusions} is the conclusions and discussion. More details about the waveforms generated by Gauss bump sources are presented in Appendix \ref{sec: waveforms_sourced_by_Gauss_bump}. Throughout this paper, we use units $c=G=M=1$.

\section{Piecewise step approximation of Regge-Wheeler potential}\label{Piecewise_step_approximation}
In black hole perturbation theory, radial axial perturbations in the time-domain on a Schwarzschild black hole are governed by the Regge-Wheeler (R-W) equation:
\begin{eqnarray}\label{Regge_Wheeler_equation}
   \Big(\frac{\partial^2}{\partial t^2}- \frac{\partial^2}{\partial x^2}+ V_{\text{R-W}}(r(x))\Big)\Psi(t,x) = 0\, ,
\end{eqnarray}
where $V_{\text{R-W}}(r)$ represents the Regge-Wheeler potential, given by
\begin{eqnarray}\label{RW_potential}
   V_{\text{R-W}}(r)=\Big(1-\frac{2M}{r}\Big)\Big[\frac{\ell(\ell+1)}{r^2}+(1-s^2)\frac{2M}{r^3}\Big]\, .
\end{eqnarray}
Here, $t$ is the coordinate time, $M$ is the mass of the black hole, and $\ell$ is the angular momentum quantum number in the spherical harmonic decomposition. The index $s$ denotes the spin of the perturbed field, with $s=0$ for the scalar, $s=1$ for electromagnetic, and $s=2$ for gravitational perturbations. The tortoise coordinate $x$ related to the radial coordinate $r$ is chosen as 
\begin{eqnarray}\label{tortoise_coordinate}
   \mathrm{d}x=\frac{\mathrm{d}r}{1-2M/r}\, ,\quad  x=r+2M\ln\Big(\frac{r-2M}{2M}\Big)\, .
\end{eqnarray}
In the following, we always concentrate on case $\ell=2$ and $s=2$. Perturbations of other parities and spins satisfy the same master equation with similar effective potentials, consequently, the qualitative conclusions drawn here should remain valid for them.

\begin{figure}[htbp]
    \centering
    \subfigure[]{\includegraphics[width=0.45\linewidth]{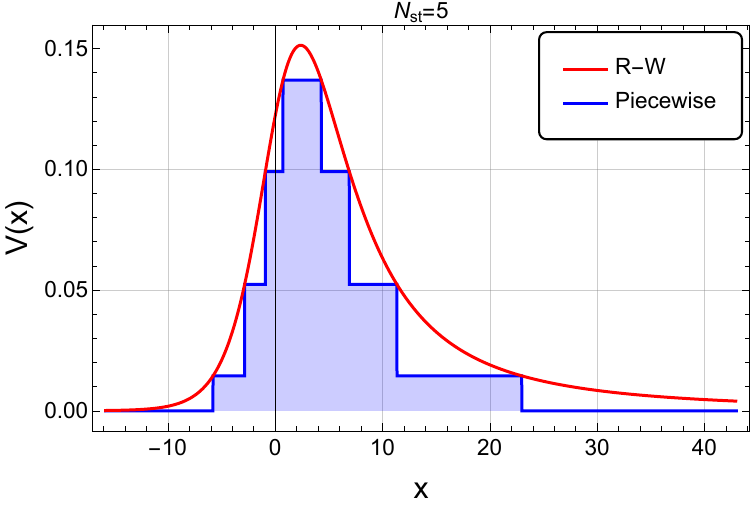}}  \hfill
    \subfigure[]{\includegraphics[width=0.45\linewidth]{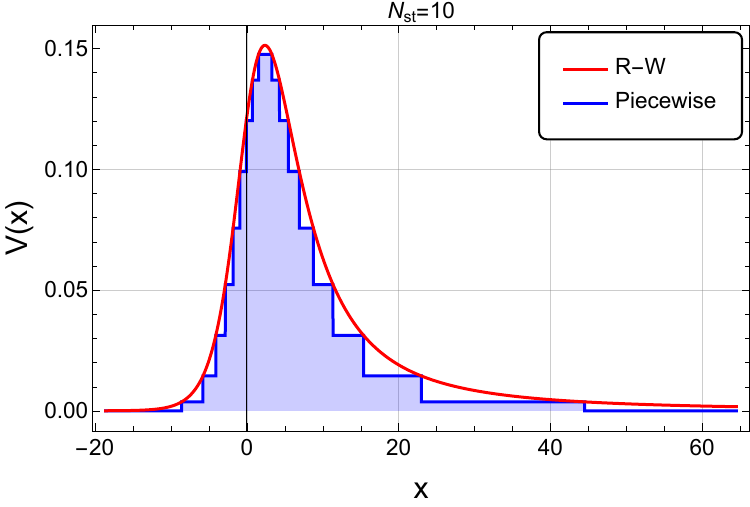}}  \\
    \subfigure[]{\includegraphics[width=0.45\linewidth]{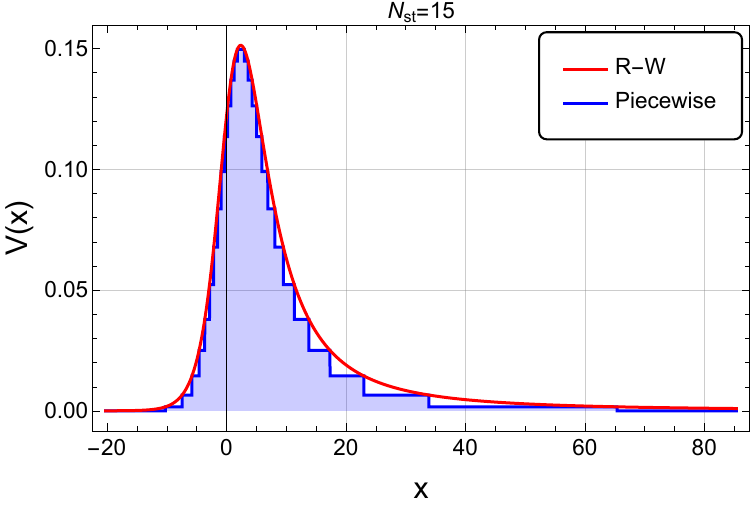}}  \hfill
    \subfigure[]{\includegraphics[width=0.45\linewidth]{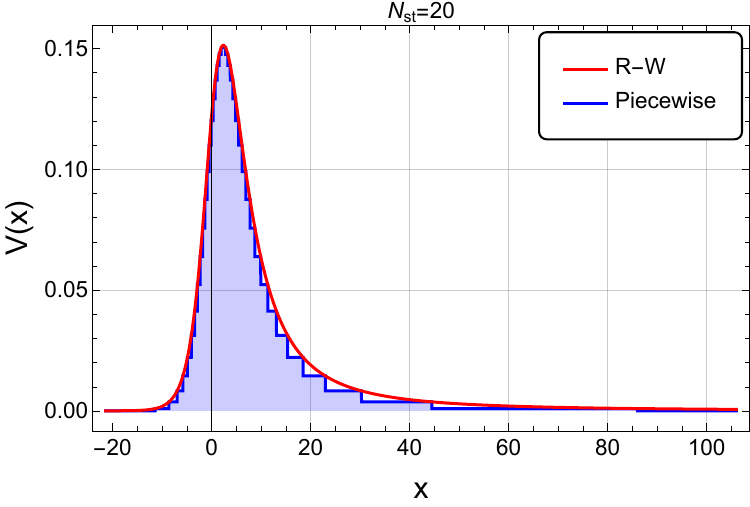}}  \\
    \subfigure[]{\includegraphics[width=0.45\linewidth]{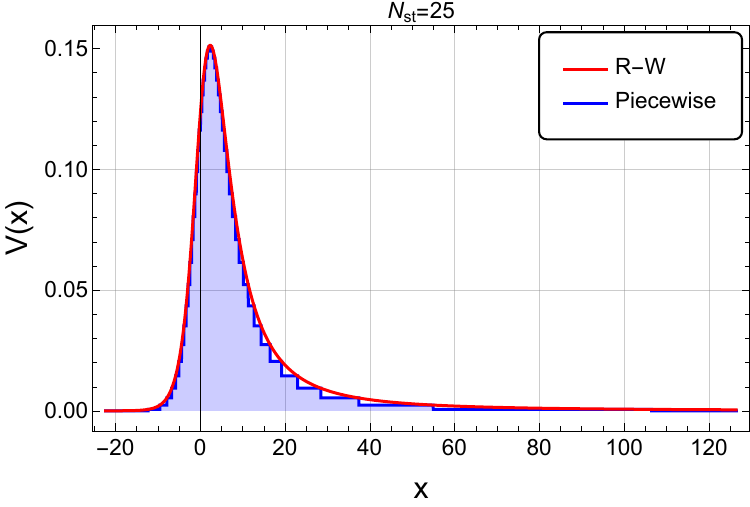}}  \hfill
    \subfigure[]{\includegraphics[width=0.45\linewidth]{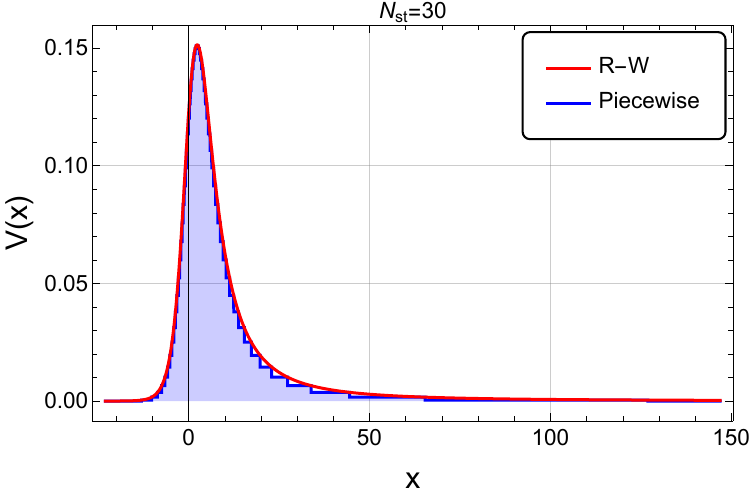}}  \\
    \caption{Comparisons between the R-W potential (red curves) and the effective potential obtained via piecewise step potential (blue curves) for $\ell=2$ and $s=2$ case with $N_{\text{st}}=5$, $10$, $15$, $20$, $25$, and $30$, respectively.}
    \label{potential_approximation}
\end{figure}

From now on, we will provide a specific strategy to achieve the piecewise step approximation of the R-W potential. Later, we will see that although the specific approximation method is different from~\cite{Nollert:1996rf}, it can still effectively achieve piecewise step approximation of the effective potential. Effective potentials $V_{\text{R-W}}$ including all $l$ and $s$ have only one extremum outside the event horizon. We denote such an extremum as $r_m$ and record the corresponding tortoise coordinate as $x_m=x(r_m)$. Here, we always make the approximation in terms of the tortoise coordinate $x$. At $x=x_m$, we divide the range $[0,V_{\text{R-W}}(x_m)]$ into $N_{\text{st}}\ge2$ parts. If horizontal lines are drawn for all points denoted by $V_0>V_1>\cdots>V_{N_{\text{st}}}$ in the interval $[0,V_{\text{R-W}}(x_m)]$ except for the first and last points, they must have two intersections with the function $V_{\text{R-W}}(x)$. We will sort the coordinates of these intersection points in order of size as $x_1,x_2,\cdots,x_{N-1},x_N$ with $N=2(N_{\text{st}}-1)$. In summary, the piecewise step approximation potential $V_{\text{st}}(x)$ is written as
\begin{eqnarray}\label{approximation_potential}
    V_{\text{st}}(x)=\left\{
    \begin{array}{l}
    0\, ,\quad x<x_1 \, ,\quad \text{or}\quad  x>x_N \, ,\\
    V_{N_{\text{st}}-1}\, ,\quad x_1<x<x_2\, ,\quad \text{or}\quad x_{N-1}<x<x_N\, ,\\
    \cdots\\
    V_{1}\, ,\quad x_{N_{\text{st}}-1}<x<x_{N_{\text{st}}}\, ,
    \end{array}\right.
\end{eqnarray}
and a series of points $V_k$ is chosen as 
\begin{eqnarray}\label{approximation_potential_Cheb}
    V_k=V_{\text{R-W}}(x_m)\Big[\frac{1+\cos(k\pi/N_{\text{st}})}{2}\Big]\, ,\quad k=1,\cdots,N_{\text{st}}-1\, .
\end{eqnarray}
The Chebyshev grid provides enhanced resolution near the prominent features of the effective potential, placing a higher density of points around the peak $x_m$ as well as in the regions close to the horizon and toward asymptotic infinity. As a typical example, in Fig. \ref{potential_approximation}, we present some approximations of the R-W potential using various numbers of $N_{\text{st}}$, with $N_{\text{st}}=5$, $10$, $15$, $20$, $25$, $30$. Intuitively, the degree of the approximation is enhanced as $N_{\text{st}}$ increases.

\section{Reviews of QNM spectra for the piecewise step potential}\label{sec: QNMs}
Although the QNM spectra for piecewise step effective potentials have already been provided in~\cite{Nollert:1996rf}, a detailed and reliable numerical method is lacking. It is still necessary to review such calculation process in order to enhance the completeness and readability of this study. Based on the above statement, in this section, we give a review on getting the QNM spectra for the piecewise step potential $V_{\text{st}}(x)$ within the transfer matrix technique~\cite{Ianniccari:2024ysv,Xie:2025jbr,Rosato:2025byu}. We are confident that this concept of transfer matrices is also applicable to multiple scenarios, such as those involving more than three isolated effective potentials. This will constitute a direction for our future research.

We investigate the problem in the frequency-domain. The Fourier component of the ﬁeld is governed by 
\begin{eqnarray}\label{Regge_Wheeler_equation_frequency_domain}
   \Big(\frac{\mathrm{d}^2}{\mathrm{d}x^2}+\omega^2- V(x)\Big)\tilde{\Psi}(\omega,x) = 0\, ,
\end{eqnarray}
with $V(x)=V_{\text{st}}(x)$. From the construction of the piecewise step effective potential in Sec. \ref{Piecewise_step_approximation}, the number of interfaces is $N$, and their positions are denoted as $x_1,\cdots,x_N$. Given $\omega\in\mathbb{C}$, we assume that the piecewise step potential $V_{\text{st}}$ is satisfied with $V_{\text{st}}(x)=V_j$, $x\in[x_{j},x_{j+1}]$, $j=0,1,\cdots,N-1,N$ where for convenience we have indicated $x_0=-\infty$, $x_{N+1}=+\infty$, and $V_{0}=V_{N}=0$. As will be readily seen, the solution in the interval $x\in[x_{j},x_{j+1}]$ can generally be expressed as
\begin{eqnarray}
    \tilde{\Psi}_{j}(x)=A_j\mathrm{e}^{-\mathrm{i}k_jx}+B_j\mathrm{e}^{\mathrm{i}k_jx}\, ,\quad j=0,\cdots,N\, .
\end{eqnarray}
Hence, the wave number $k_j$ in each interval $[x_j,x_{j+1}]$ is given by $k_j=\sqrt{\omega^2-V_j}$. Note that $k_0=k_N=\omega$, and $k_j=\sqrt{\omega^2-V_j}$, $j=1,\cdots,N-1$. We call $[A_j,B_j]^{\text{T}}$ the state vector. For each interface $x=x_j$, using the continuity conditions of the solutions and their derivatives, we have
\begin{eqnarray}
    A_{j-1}\mathrm{e}^{-\mathrm{i}k_{j-1}x_j}+B_{j-1}\mathrm{e}^{\mathrm{i}k_{j-1}x_j}&=&A_j\mathrm{e}^{-\mathrm{i}k_jx_j}+B_j\mathrm{e}^{\mathrm{i}k_jx_j}\, ,\nonumber\\
    A_{j-1}(-\mathrm{i}k_{j-1})\mathrm{e}^{-\mathrm{i}k_{j-1}x_j}+B_{j-1}(\mathrm{i}k_{j-1})\mathrm{e}^{\mathrm{i}k_{j-1}x_j}&=&A_j(-\mathrm{i}k_j)\mathrm{e}^{-\mathrm{i}k_jx_j}+B_j(\mathrm{i}k_j)\mathrm{e}^{\mathrm{i}k_jx_j}\, .
\end{eqnarray}
We can write the above equation to a matrix equation, i.e., 
\begin{eqnarray}
    \begin{bmatrix}
        A_j\\
        B_j
    \end{bmatrix}=
    \begin{bmatrix}
        \mathrm{e}^{-\mathrm{i}k_jx_j} & \mathrm{e}^{\mathrm{i}k_jx_j}\\
        (-\mathrm{i}k_j)\mathrm{e}^{-\mathrm{i}k_jx_j} & (\mathrm{i}k_j)\mathrm{e}^{\mathrm{i}k_jx_j}
    \end{bmatrix}^{-1}
    \bm{\cdot}
    \begin{bmatrix}
        \mathrm{e}^{-\mathrm{i}k_{j-1}x_j} & \mathrm{e}^{\mathrm{i}k_{j-1}x_j}\\
        (-\mathrm{i}k_{j-1})\mathrm{e}^{-\mathrm{i}k_{j-1}x_j} & (\mathrm{i} k_{j-1})\mathrm{e}^{\mathrm{i}k_{j-1}x_j} 
    \end{bmatrix}
    \bm{\cdot}
    \begin{bmatrix}
        A_{j-1}\\
        B_{j-1}
    \end{bmatrix}\, .
\end{eqnarray}
This is a recurrence relation from which we have
\begin{eqnarray}\label{A_n_and_B_n}
    \begin{bmatrix}
        A_{n}\\
        B_{n}
    \end{bmatrix}=\Big(\prod_{j=1}^{n}\mathbf{T}[j]\Big)\bm{\cdot}
    \begin{bmatrix}
        A_{0}\\
        B_{0}
    \end{bmatrix}\, ,\quad n=1,2,\cdots,N-1,N\, ,
\end{eqnarray}
where
\begin{eqnarray}\label{T_matrix}
    \mathbf{T}[j]=
    \begin{bmatrix}
        \mathrm{e}^{-\mathrm{i}k_jx_j} & \mathrm{e}^{\mathrm{i}k_jx_j}\\
        (-\mathrm{i}k_j)\mathrm{e}^{-\mathrm{i}k_jx_j} & (\mathrm{i}k_j)\mathrm{e}^{\mathrm{i}k_jx_j}
    \end{bmatrix}^{-1}
    \bm{\cdot}
    \begin{bmatrix}
        \mathrm{e}^{-\mathrm{i}k_{j-1}x_j} & \mathrm{e}^{\mathrm{i}k_{j-1}x_j}\\
        (-\mathrm{i}k_{j-1})\mathrm{e}^{-\mathrm{i}k_{j-1}x_j} & (\mathrm{i}k_{j-1})\mathrm{e}^{\mathrm{i}k_{j-1}x_j} 
    \end{bmatrix}\, .
\end{eqnarray}
The above matrix $\mathbf{T}[j]$ is called the transfer matrix, which is a key quantity for our present study. Note that states are transferred from the left side to the right side. We focus on the ``in'' solution satisfying the purely ingoing boundary condition at the horizon, consequently exhibiting asymptotic behavior characterized by
\begin{eqnarray}\label{in_solution}
    \tilde{\Psi}_{\text{in}}(x)=
    \left\{
    \begin{array}{l}
    \mathrm{e}^{-\mathrm{i}\omega x}\, ,\quad x\to-\infty\\
    A_{\text{in}}(\omega)\mathrm{e}^{-\mathrm{i}\omega x}+A_{\text{out}}(\omega)\mathrm{e}^{\mathrm{i}\omega x}\, ,\quad x\to+\infty
    \end{array}\right.\, .
\end{eqnarray}
One can use the transfer matrix technique to solve $A_{\text{in}}(\omega)$ and $A_{\text{out}}(\omega)$ analytically. In the language of the transfer matrix mentioned above, given $A_0=1$, $B_0=0$, we can obtain $A_{\text{in}}=A_N$, $A_{\text{out}}=B_N$ from Eq. (\ref{A_n_and_B_n}) with the potential given by Eq. (\ref{approximation_potential}) and Eq. (\ref{approximation_potential_Cheb}). It is worth mentioning that the double square barrier potential, as the simplest example of studying the effect of ``flea on elephant'' on spectrum instability, can be directly calculated using the transfer matrix formalism with $N=4$, and $V_2=0$~\cite{Cheung:2021bol,Kyutoku:2022gbr,Barausse:2014tra,Motohashi:2024fwt,Qian:2024iaq}. For $N=2$ case, i.e., a single rectangular barrier, one can refer to the work of Chandrasekhar and Detweiler~\cite{Chandrasekhar:1975zza}. The ``up'' solution has the property $\tilde{\Psi}_{\text{up}}(x)=\mathrm{e}^{\mathrm{i}\omega x}$, as $x\to+\infty$. The Wronskian of these two corresponding homogeneous solutions is
\begin{eqnarray}\label{Wronskian}
    W=2\mathrm{i}\omega A_{\text{in}}(\omega)\, .
\end{eqnarray}
QNM spectra are defined by the requirement of purely ingoing waves at the event horizon and purely outgoing waves at asymptotic infinity. This specific boundary condition implies that, for a QNM spectrum, the global ``in'' and ``up'' solutions are linearly dependent. It follows that the Wronskian formed from these solutions vanishes at the spectra. Equivalently, these spectra correspond to the zeros of the coefficient function $A_{\text{in}}(\omega)$ within the radial solution. There is an infinity of QNM spectra, which are usually sorted by the magnitude of their imaginary part and labeled by an integer $n$ from zero. As $N_{\text{st}}$ increases, the complexity of function $A_{\text{in}}(\omega)$ also increases. The implementation for finding zeros of $A_{\text{in}}(\omega)$ is based on the Muller method, which can be found in Chapter 9.2 of~\cite{Press:2007ipz}. 

For the Muller method, the iterative process commences with three initial values. If these initial selections are substantially distant from the true QNM spectra, the convergence of the Muller method may proceed at a significantly reduced rate or, in certain cases, fail to converge entirely. For the case of adding a single small bump, where the parameters are continuously changing, the QNM spectra of the previous iteration can serve as the initial values for the next iteration. However, our discrete parameter $N_{\text{st}}$ gives rise to marked disparities in the distribution of QNM spectra. Based on the above, we propose a method to accurately find the appropriate initial iteration values in the Muller method, and thus find all QNM spectra within some specified region. In fact, since we know the analytical expression of $A_{\text{in}}(\omega)$, we discretize a certain region of the $\omega$ complex plane of interest using grid points, then calculate the value of $A_{\text{in}}(\omega)$ and $\ln|A_{\text{in}}(\omega)|$ at each grid point. Finally, the local minimum of the obtained two-dimensional discrete matrix $\ln|A_{\text{in}}(\Re(\omega)+\mathrm{i}\Im(\omega))|$ becomes the candidate of the initial iteration value in the Muller method. By doing so, the convergence of the Muller method will be greatly enhanced. In \textit{Mathematica}, one can use the built-in function \textit{MinDetect} to achieve finding local minimum.

\begin{figure}[htbp]
	\centering
	\includegraphics[width=0.8\textwidth]{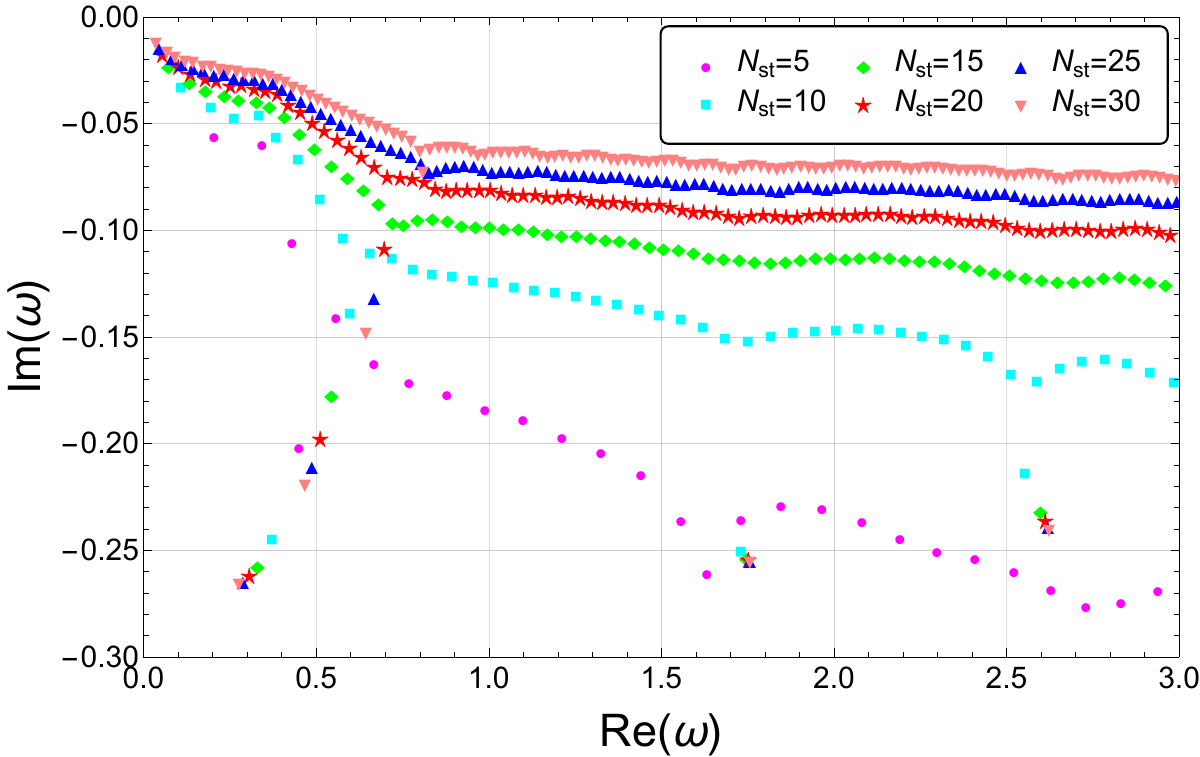}
	\caption{For $\ell=2$ and $s=2$, QNM spectra with various $N_{\text{st}}$ are displayed. Magenta $\bullet$ stands for $N_{\text{st}}=5$, cyan $\blacksquare$ stands for $N_{\text{st}}=10$, green $\blacklozenge$ stands for $N_{\text{st}}=15$, red $\star$ stands for $N_{\text{st}}=20$, blue $\blacktriangle$ stands for $N_{\text{st}}=25$, and black $\blacktriangledown$ stands for $N_{\text{st}}=30$. The displayed range for the QNM spectra is $[0,3]\times[-0.3,0]$. In addition, the discrimination criterion of our Muller method is $|A_{\text{in}}|<10^{-100}$.} 
	\label{QNM_spectra}
\end{figure}

Using the iterative initial value search method given above, we can calculate the QNM spectra for a given $N_{\text{st}}$. The results of QNM spectra with $N_{\text{st}}=5$, $10$, $15$, $20$, $25$, $30$ are shown in Fig. \ref{QNM_spectra}. Among them, we do not find any pure imaginary modes, so based on spherical symmetry, we only show modes with real parts greater than $0$. As shown in Fig. \ref{QNM_spectra}, unlike the QNM spectra of Schwarzschild case, the QNM spectra associated with the piecewise step potentials exhibit a clear bifurcation behavior: one dominant (dense) branch lies along the real axis, while the others are sparsely distributed across the complex plane. For the (dense) branch, the imaginary parts $(-\operatorname{Im}(\omega))$ of these QNM spectra, are generally small and grow slowly with increasing overtone number. Furthermore, our results indicate that as the number of steps increases, the branch of QNM spectra distributed along the real axis tends to lie closer to the axis, and the QNM spectra become more densely packed. These results are consistent with~\cite{Nollert:1996rf}.

\section{Excitation factors for the piecewise step potential}\label{sec: EFs}
A black hole possesses the infinite number of discrete QNM spectra. The resultant ringdown waveform is fundamentally constituted as a superposition of these modes. Crucially, excitation amplitudes exhibit significant variability across the spectrum contingent upon the perturbation source. This necessitates a quantitative framework to characterize the excitation susceptibility intrinsic to each QNM. This quantity is exactly what is called the excitation factors (EFs)~\cite{Leaver:1986gd,Sun:1988tz,Andersson:1995zk,Glampedakis:2003dn,Berti:2006wq,Zhang:2013ksa,Silva:2024ffz,Oshita:2024wgt,Oshita:2025ibu,Lo:2025njp,Takahashi:2025uwo,Oshita:2021iyn,Chen:2024hum,Motohashi:2024fwt,Kubota:2025hjk}. These complex-valued constants constitute intrinsic attributes of a black-hole spacetime's perturbation spectrum, fundamentally governing the relative amplitude with which distinct QNMs are excited in response to a given initial perturbation profile. It is known that the integration constant of the tortoise coordinate introduces an ambiguity in the EFs. One can refer to~\cite{Kubota:2025hjk}, in which such ambiguity is clarified and explained in detail. Having delineated the method for calculating the QNM spectra, we herein characterize the derivation of their corresponding EFs for the piecewise step potential, which have never been studied anywhere as far as we know. For conciseness, we exclusively delve into the excitation factors of spherically symmetric black holes and revisit foundational concepts relevant to their analysis. This approach simultaneously serves the dual purpose of laying the groundwork for subsequent waveform stability discussions.

In this context, we take an interest in the inhomogeneous version of Eq. (\ref{Regge_Wheeler_equation_frequency_domain}) becomes 
\begin{eqnarray}\label{RW_equation_frequency_domain_inhomogeneous}
    \frac{\mathrm{d}^2\tilde{\Psi}(\omega,x)}{\mathrm{d}x^2}+\Big[\omega^2-V_{\text{st}}(x)\Big]\tilde{\Psi}(\omega,x)=I(\omega,x)\, ,
\end{eqnarray}
where $\tilde{\Psi}(\omega,x)$ is the Laplace transformation of $\Psi(t,x)$. The function $I(\omega,x)$ is called the source term, which comes from the initial condition via
\begin{eqnarray}
    I(\omega,x)=\Big[\mathrm{i}\omega\Psi(t,x)-\frac{\partial\Psi(t,x)}{\partial t}\Big]\Big|_{t=0}\, .
\end{eqnarray}
Using the Green's function method, the general solution of Eq. (\ref{RW_equation_frequency_domain_inhomogeneous}) is expressed as~\cite{Andersson:1995zk,Berti:2006wq,Zhang:2013ksa,Berti:2025hly}
\begin{eqnarray}\label{general_solution}
    \tilde{\Psi}(\omega,x)=\tilde{\Psi}_{\text{up}}(\omega,x)\int_{-\infty}^{x}\frac{I(\omega,x^{\prime})\tilde{\Psi}_{\text{in}}(\omega,x^{\prime})}{2\mathrm{i}\omega A_{\text{in}}(\omega)}\mathrm{d}x^{\prime}+\tilde{\Psi}_{\text{in}}(\omega,x)\int_{x}^{+\infty}\frac{I(\omega,x^{\prime})\tilde{\Psi}_{\text{up}}(\omega,x^{\prime})}{2\mathrm{i}\omega A_{\text{in}}(\omega)}\mathrm{d}x^{\prime}\, .
\end{eqnarray}
Under standard astrophysical considerations, we assume an observer at considerable remove from the black hole to proceed with the initial data possess compact support. Then the solution (\ref{general_solution}) has a fine approximation in the frequency-domain, which is given by~\cite{Oshita:2024wgt}
\begin{eqnarray}\label{general_solution_far_observer}
    \tilde{\Psi}(\omega,x)=\mathrm{e}^{\mathrm{i}\omega x}\tilde{\Psi}_{\text{G}}(\omega)\tilde{\Psi}_{\text{T}}(\omega)\, ,\quad \text{as} \quad x\to+\infty\, ,
\end{eqnarray}
where the expressions of two functions $\tilde{\Psi}_{\text{G}}(\omega)$ and $\tilde{\Psi}_{\text{T}}(\omega)$ are
\begin{eqnarray}\label{tilde_Psi_G_and_tilde_Psi_T}
    \tilde{\Psi}_{\text{G}}(\omega)\equiv\frac{A_{\text{out}}(\omega)}{2\mathrm{i}\omega A_{\text{in}}(\omega)}\, ,\quad \tilde{\Psi}_{\text{T}}(\omega)=\int_{-\infty}^{+\infty}\frac{I(\omega,x^{\prime})\tilde{\Psi}_{\text{in}}(\omega,x^{\prime})}{A_{\text{out}}(\omega)}\mathrm{d}x^{\prime}\, . 
\end{eqnarray}
In Eq. (\ref{general_solution_far_observer}), the inverse transformation of the factor $\tilde{\Psi}_{\text{G}}(\omega)$ is called the fundamental ringdown denoted by $\Psi_\text{G}(t,x)$ which is given by~\cite{Oshita:2024wgt,Oshita:2025ibu}
\begin{eqnarray}\label{fundamental_ringdown}
    \Psi_\text{G}(t,x)=\frac{1}{2\pi}\int_{-\infty}^{+\infty}\mathrm{d}\omega \tilde{\Psi}_{\text{G}}(\omega)\mathrm{e}^{-\mathrm{i}\omega (t-x)}\, .
\end{eqnarray}
This waveform is obviously unrelated to the source. An actual ringdown waveform comes from the convolution of the fundamental ringdown $\Psi_{\text{G}}(t,x)$ with the waveform $\Psi_{\text{T}}(t,x)$, which is defined as the inverse transform of $\tilde{\Psi}_{\text{T}}(\omega)$. where the function $\tilde{\Psi}_{\text{T}}(\omega)$ associated with $\omega$ is called the source factor~\cite{Oshita:2021iyn}. This quantity includes the information of the source of perturbation. Hence, in the time-domain, the specific form of the solution is
\begin{eqnarray}\label{general_solution_time_domain}
    \Psi(t,x)=\frac{1}{2\pi}\int_{-\infty}^{+\infty}\mathrm{d}\omega\tilde{\Psi}(\omega,x) \mathrm{e}^{-\mathrm{i}\omega t}\, .
\end{eqnarray}
Instead of performing the above integral along the real line, we can compute it by performing a contour integral in the complex plane. For the contour integral, one can refer to the Fig. 2 in~\cite{Berti:2006wq}. According to the residue theorem, the contour integral reduces to a discrete sum over the residues evaluated at the poles of the integrand, i.e.,  all QNM spectra $\omega_n$ included in the contour integral. Therefore, in the time-domain the solution $\Psi(t,x)$ can be expanded into a sum associated with QNMs, which has the following form
\begin{eqnarray}\label{general_solution_time_domain_sum_QNMs}
    \Psi(t,x)=-\sum_{\omega_n}E(\omega_n)\tilde{\Psi}_{\text{T}}(\omega_n)\mathrm{e}^{-\mathrm{i}\omega_n(t-x)}\, ,
\end{eqnarray}
where the part $E(\omega_n)$ that is independent of the source part is called the excitation factor (EF). Its definition is
\begin{eqnarray}\label{EFs}
    E(\omega_n)\equiv\frac{A_{\text{out}}}{2\omega}\Big(\frac{\mathrm{d}A_{\text{in}}}{\mathrm{d}\omega}\Big)^{-1}\Bigg|_{\omega=\omega_n}\, .
\end{eqnarray}
Note that the minus sign before $\Sigma$ in Eq. (\ref{general_solution_time_domain_sum_QNMs}) is due to the fact that the contour integral takes the lower half plane (The poles of Green's functions occupy the lower half of the complex plane.). 

\begin{figure}[htbp]
    \centering
    \subfigure[]{\includegraphics[width=0.45\linewidth]{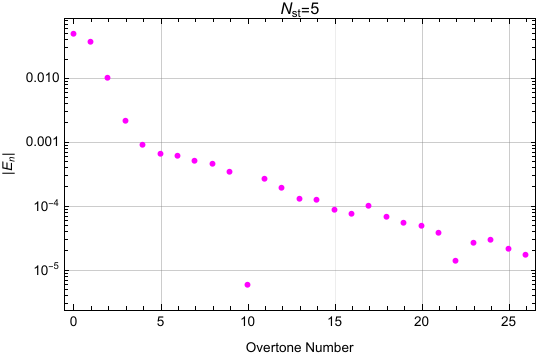}}  \hfill
    \subfigure[]{\includegraphics[width=0.45\linewidth]{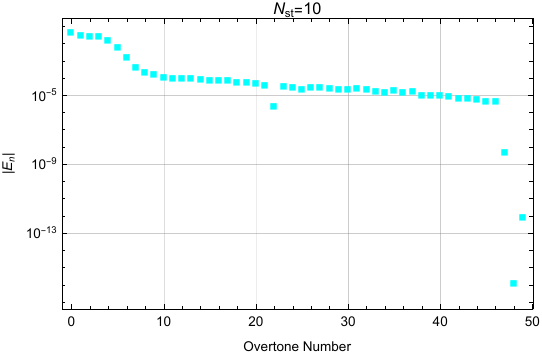}}  \\
    \subfigure[]{\includegraphics[width=0.45\linewidth]{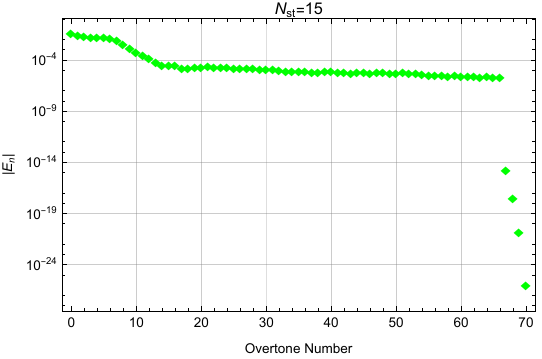}}  \hfill
    \subfigure[]{\includegraphics[width=0.45\linewidth]{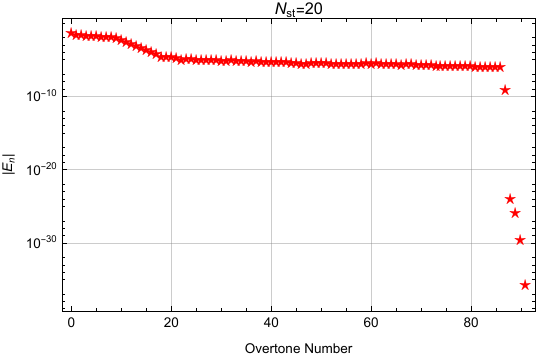}}  \\
    \subfigure[]{\includegraphics[width=0.45\linewidth]{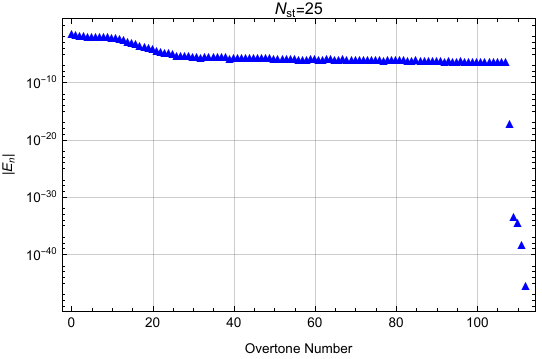}}  \hfill
    \subfigure[]{\includegraphics[width=0.45\linewidth]{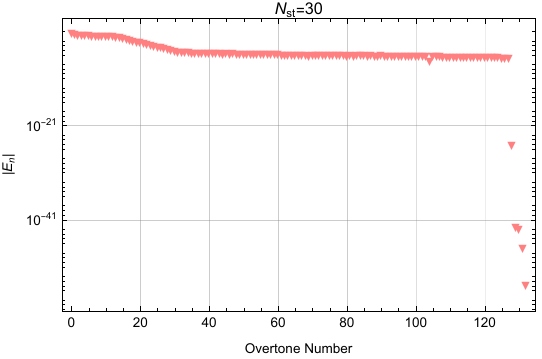}}  \\
    \caption{The absolutes of the QNM EFs with various $N_{\text{st}}$ are displayed, where the horizontal axis is the overtone number. Overtones are sorted by the absolutes of the imaginary parts of spectra. For each panels, magenta $\bullet$ stands for $N_{\text{st}}=5$, cyan $\blacksquare$ stands for $N_{\text{st}}=10$, green $\blacklozenge$ stands for $N_{\text{st}}=15$, red $\star$ stands for $N_{\text{st}}=20$, blue $\blacktriangle$ stands for $N_{\text{st}}=25$, and pink $\blacktriangledown$ stands for $N_{\text{st}}=30$.}
    \label{abs_EFs_piecewise}
\end{figure}

In Sec. \ref{sec: QNMs}, for the piecewise step potential $V_{\text{st}}(x)$, we have obtained the expressions of $A_{\text{in}}(\omega)$ and $A_{\text{out}}(\omega)$ using the transfer matrix technique. So the excitation factors are able to be calculated directly. The results for the absolutes of the QNM EFs with $N_{\text{st}}=5$, $10$, $15$, $20$, $25$, $30$ are shown in Fig. \ref{abs_EFs_piecewise}. In practice, we use the sixth-order central difference formula to calculate the derivative of $A_{\text{in}}(\omega)$, where the step size $h=10^{-20}$ is used. The displayed excitation factors correspond to the QNM spectra shown in Fig. \ref{QNM_spectra}, and we have arranged the QNM spectra according to the magnitude of the imaginary part, with the excitation factors correspondingly aligned. In Fig. \ref{abs_EFs_piecewise}, these inconsistent points correspond exactly to those points in Fig. \ref{QNM_spectra} that are not on the main branch of QNM spectra. The magnitude of the excitation factor generally decreases with increasing overtone number, which also reflects the fact that, for an individual mode, the larger the absolute value of its imaginary part, the lower the degree of excitation. Moreover, the reason why the excitation factors corresponding to the sparsely distributed QNM spectra are so small can also be inferred from the fact: these spectra are located far from their neighboring modes, which results in a large derivative of $A_{\text{in}}(\omega)$ at these points, and consequently, according to Eq. (\ref{EFs}), the excitation factors are suppressed.

\section{Stability of waveforms}\label{sec: stability_of_waveforms}
In Ref. \cite{Oshita:2025ibu}, the authors study the ringdown waveform sourced by a Dirac delta function, which demonstrates the prompt time-domain signals are robust in the case of adding a single bump to the original potential. Although they changed the boundary conditions of QNMs, such modification effectively serves as a remotely added bump to the effective potential. How will the piecewise step approximation of the effective potential affect the waveform? In essence, approximating the effective potential via discrete steps is equivalent to introducing numerous tightly clustered perturbations distributed across the original potential rather than a single one. In this section, we consider the stability of the waveforms from the piecewise step potential by comparing them with the corresponding Schwarzschild waveforms. All waveforms presented here are the solutions of Eq. (\ref{RW_equation_frequency_domain_inhomogeneous}). For simplicity, we restrict our consideration exclusively to scenarios where waveforms constructed from far-distance and near-horizon source.

We start from considering two waveforms with the following forms~\cite{Oshita:2024wgt,Oshita:2025ibu,Rosato:2025byu,Rosato:2024arw}:
\begin{eqnarray}\label{fundamental_ringdown_H}
    \Psi_{\text{H}}(t,x)\equiv\Psi_{\text{H}}(u)=\frac{1}{2\pi}\int_{-\infty}^{+\infty}\mathrm{d}\omega H(\omega) \mathrm{e}^{-\mathrm{i}\omega u}\, ,\quad H(\omega)\equiv\frac{A_{\text{out}}(\omega)}{A_{\text{in}}(\omega)}\, ,
\end{eqnarray}
and
\begin{eqnarray}\label{fundamental_ringdown_F}
    \Psi_{\text{F}}(t,x)\equiv\Psi_{\text{F}}(u)=\frac{1}{2\pi}\int_{-\infty}^{+\infty}\mathrm{d}\omega F(\omega) \mathrm{e}^{-\mathrm{i}\omega u}\, ,\quad F(\omega)\equiv\frac{1}{A_{\text{in}}(\omega)}\, ,
\end{eqnarray}
where we have defined the retarded time $u=t-x$. The function $H(\omega)$ represents the reflection amplitude while the function $F(\omega)$ represents the transmission amplitude. Furthermore, the relationship between reflection coefficient and reflection amplitude, as well as transmission coefficient (graybody factor) and transmission amplitude are
\begin{eqnarray}\label{reflection_amplitude_reflection_coefficient}
    R(\omega)=|H(\omega)|^2\, ,\quad \text{with}\quad  H(\omega)=|H(\omega)|\mathrm{e}^{\mathrm{i}\delta (\omega)}\, ,
\end{eqnarray}
and
\begin{eqnarray}\label{reflection_amplitude_reflection_coefficient}
    \Gamma(\omega)=|F(\omega)|^2\, ,\quad \text{with}\quad  F(\omega)=|F(\omega)|\mathrm{e}^{\mathrm{i}\eta (\omega)}\, ,
\end{eqnarray}
where the phases $\delta(\omega)$ of $H(\omega)$ and $\eta(\omega)$ of $F(\omega)$ have been explicitly written out. Compared to QNM spectra, graybody factors possess superior robustness against slight external perturbations, conferring greater stability that has been the focus of considerable studies~\cite{Oshita:2024fzf,Wu:2024ldo,Rosato:2024arw,Ianniccari:2024ysv,Konoplya:2025ixm,Xie:2025jbr}. In Refs. \cite{Oshita:2024fzf,Wu:2024ldo,Rosato:2024arw,Ianniccari:2024ysv,Konoplya:2025ixm}, a bump is added to the original effective potential, and the graybody factors are stable for such approach. Due to the energy conservation, i.e., $R(\omega)+\Gamma(\omega)=1$, one can also assert that the reflection coefficient $R(\omega)$ is stable. However, this merely demonstrates the stability of the magnitudes of $H(\omega)$ and $F(\omega)$, and does not necessarily imply stability in $H(\omega)$ and $F(\omega)$ themselves. In the following aspects, the stability behaviors of four functions $|H(\omega)|$, $\delta(\omega)$, $|F(\omega)|$, and $\eta(\omega)$ will be of concern~\cite{Kyutoku:2022gbr}. Note that recently, when a black hole surrounded by a thin shell of matter, the stability property of the function $F(\omega)$ or $A_{\text{in}}(\omega)$ is studied in~\cite{Laeuger:2025zgb}.


Now we come back to the Schwarzschild black hole whose potential is given by Eq. (\ref{RW_potential}). The functions $H(\omega)$ and $F(\omega)$ for real-frequency $\omega$ can be derived numerically by directly integrating $\tilde{\Psi}_{\text{in}}(\omega,x)$ with respect to $x$ from $-\infty$ to $+\infty$ along the real axis. By the way, if $\omega$ has an imaginary part, then the integration path needs to be deformed~\cite{Glampedakis:2003dn,Silva:2024ffz,Lo:2025njp}. To mitigate the influence of finite boundary effects on numerical results, we thus adopt higher-order approximations for initial conditions, which is similar to~\cite{Laeuger:2025zgb}. At the event horizon, we suppose that the ``in'' solution $\tilde{\Psi}_{\text{in}}(\omega,x)$ has the following expansion
\begin{eqnarray}\label{solution_expansion}
    \tilde{\Psi}_{\text{in}}(\omega,x)=\mathrm{e}^{-\mathrm{i}\omega x}\sum_{n=0}^{\infty}a_n\mathrm{e}^{nx/(2M)}\, ,\quad x\to-\infty\, .
\end{eqnarray}
The asymptotic behavior of the R-W potential as $x\to-\infty$ is~\cite{Daghigh:2020jyk}
\begin{eqnarray}\label{potential_event_horizon}
    V_{\text{R-W}}(x)\sim \frac{c}{4M^2}\mathrm{e}^{x/(2M)}\, ,\quad c\equiv \frac{\ell(\ell+1)+(1-s^2)}{\mathrm{e}}\, .
\end{eqnarray}
Substituting Eq. (\ref{solution_expansion}) and Eq. (\ref{potential_event_horizon}) into Eq. (\ref{Regge_Wheeler_equation_frequency_domain}), one gets
\begin{eqnarray}
    a_1=\frac{c}{1-4\mathrm{i}\omega M}a_0\, .
\end{eqnarray}
Set $a_0=1$, the ``in'' solution $\tilde{\Psi}_{\text{in}}(\omega,x)$ and its derivative have the following expansions:
\begin{eqnarray}\label{Psi_in_expansion}
    \tilde{\Psi}_{\text{in}}(\omega,x)=\mathrm{e}^{-\mathrm{i}\omega x}\Big[1+\frac{c}{1-4\mathrm{i}\omega M}\mathrm{e}^{x/(2M)}+\cdots\Big]\, ,
\end{eqnarray}
and 
\begin{eqnarray}\label{DPsi_in_expansion}
    \frac{\mathrm{d}\tilde{\Psi}_{\text{in}}(\omega,x)}{\mathrm{d}x}=\mathrm{e}^{-\mathrm{i}\omega x}\Big[-\mathrm{i}\omega+\frac{(1-2\mathrm{i}\omega M)c}{2M(1-4\mathrm{i}\omega M)}\mathrm{e}^{x/(2M)}+\cdots\Big]\, .
\end{eqnarray}
The initial conditions of $\tilde{\Psi}_{\text{in}}(\omega,x)$ and $\mathrm{d}\tilde{\Psi}_{\text{in}}(\omega,x)/\mathrm{d}x$ for finite values $x_{\text{I}}$ are chosen to truncate Eq. (\ref{Psi_in_expansion}) and Eq. (\ref{DPsi_in_expansion}) to the secondary dominant term, in which $x_{\text{I}}=-100$ in our practice. Finally, we are able to obtain the values of two functions $A_{\text{in}}(\omega)$ and $A_{\text{out}}(\omega)$ on the real axis by direct integration. Due to limitations in computational precision, we only calculate the range from $0$ to $1.5$ of $\omega$ for functions $H(\omega)$ and $F(\omega)$ of the Schwarzschild black hole with $\Delta\omega=0.0005$. As for $\omega\in[-1.5,0]$, one can easily obtain $H(\omega)$ and $F(\omega)$ by using the property that $\overline{A_{\text{in}}(\omega)}=A_{\text{in}}(-\omega)$ and $\overline{A_{\text{out}}(\omega)}=A_{\text{out}}(-\omega)$, where bar means the complex conjugate and same below.

We first concentrate on the waveform (\ref{fundamental_ringdown_H}), where the integrand is the reflection amplitude $H(\omega)$, which is also studied in~\cite{Oshita:2024wgt,Oshita:2025ibu,Rosato:2025byu,Rosato:2024arw}. Actually, this waveform can be derived from a far-distant Dirac delta source $I=2\mathrm{i}\omega\delta(x-x_s)$ with $x_s\gg x_m$, and details can be found in Appendix \ref{sec: waveforms_sourced_by_Gauss_bump} [see Eq. (\ref{waveform_sourced_delta_infinity}) and note that the amplitude $A$ is always set to unity]. At present, we compare the results of $H(\omega)$ from the piecewise step effective potential against the Schwarzschild black hole R-W potential. We can conduct comprehensive comparisons, including the magnitude of $H(\omega)$, the phase $\delta(\omega)$, and the corresponding waveform $\Psi_{\text{H}}(u)$, namely the inverse transform of $H(\omega)$. 

\begin{figure}[htbp]
	\centering
    \subfigure[]
	{\includegraphics[width=0.45\textwidth]{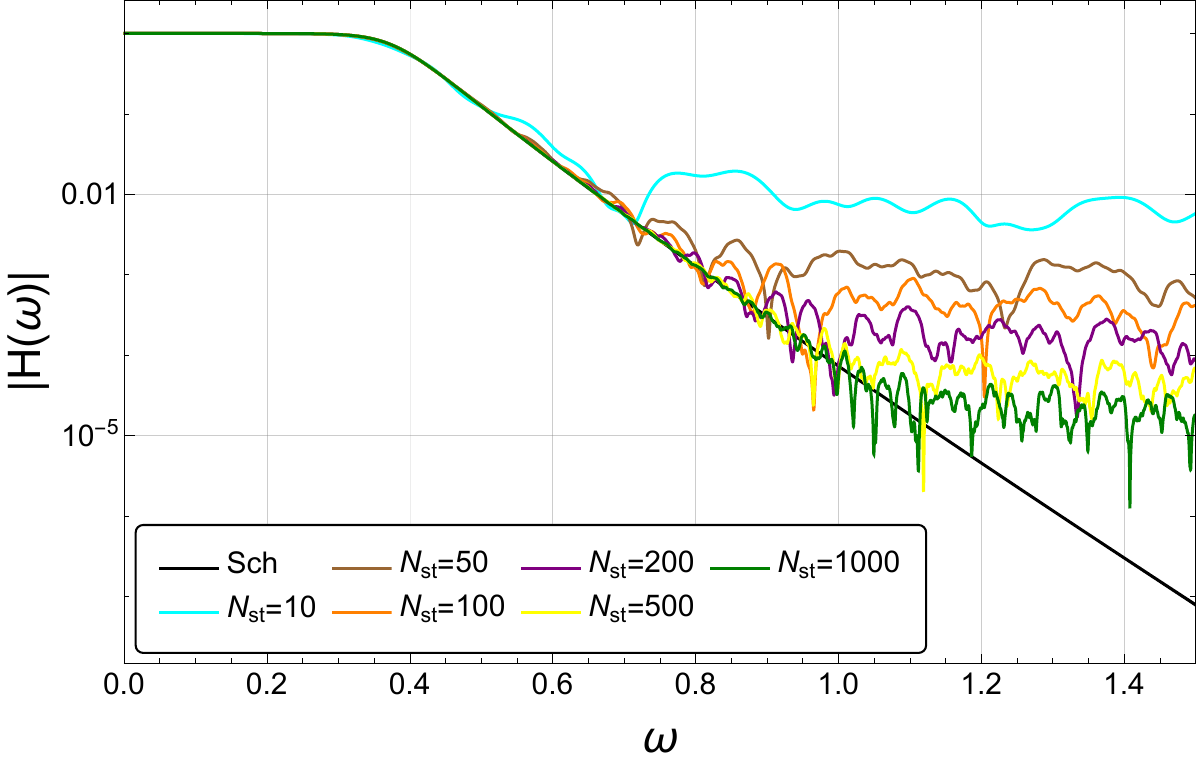}\label{Abs_H_omega}} \hfill
    \subfigure[]
    {\includegraphics[width=0.45\textwidth]{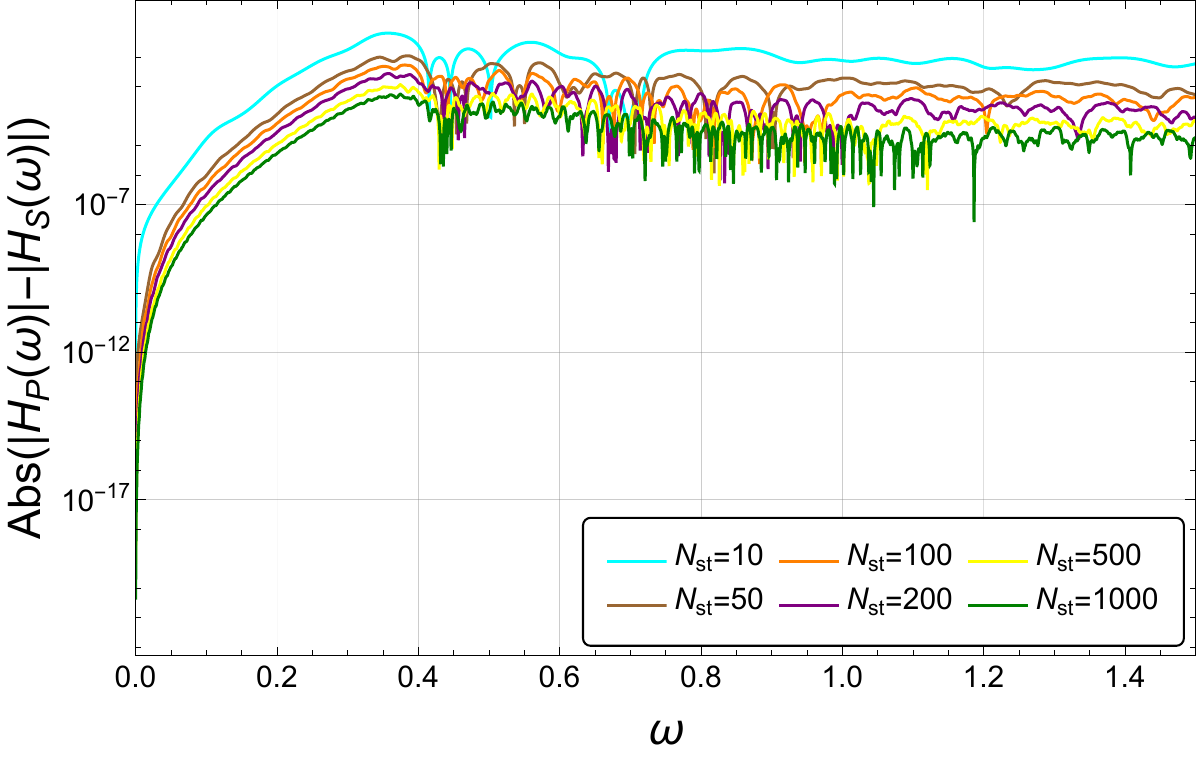}}
    \subfigure[]
    {{\includegraphics[width=0.45\linewidth]{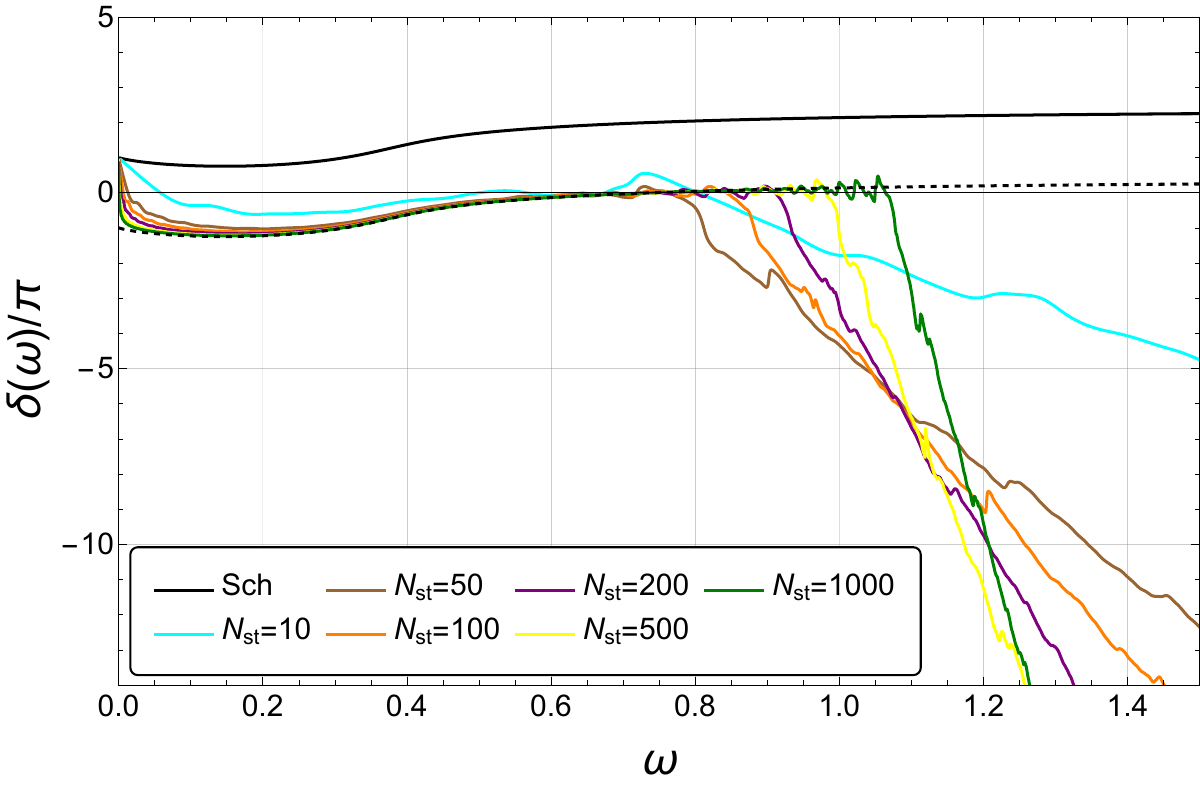}}} \hfill
    \subfigure[]
    {{\includegraphics[width=0.45\linewidth]{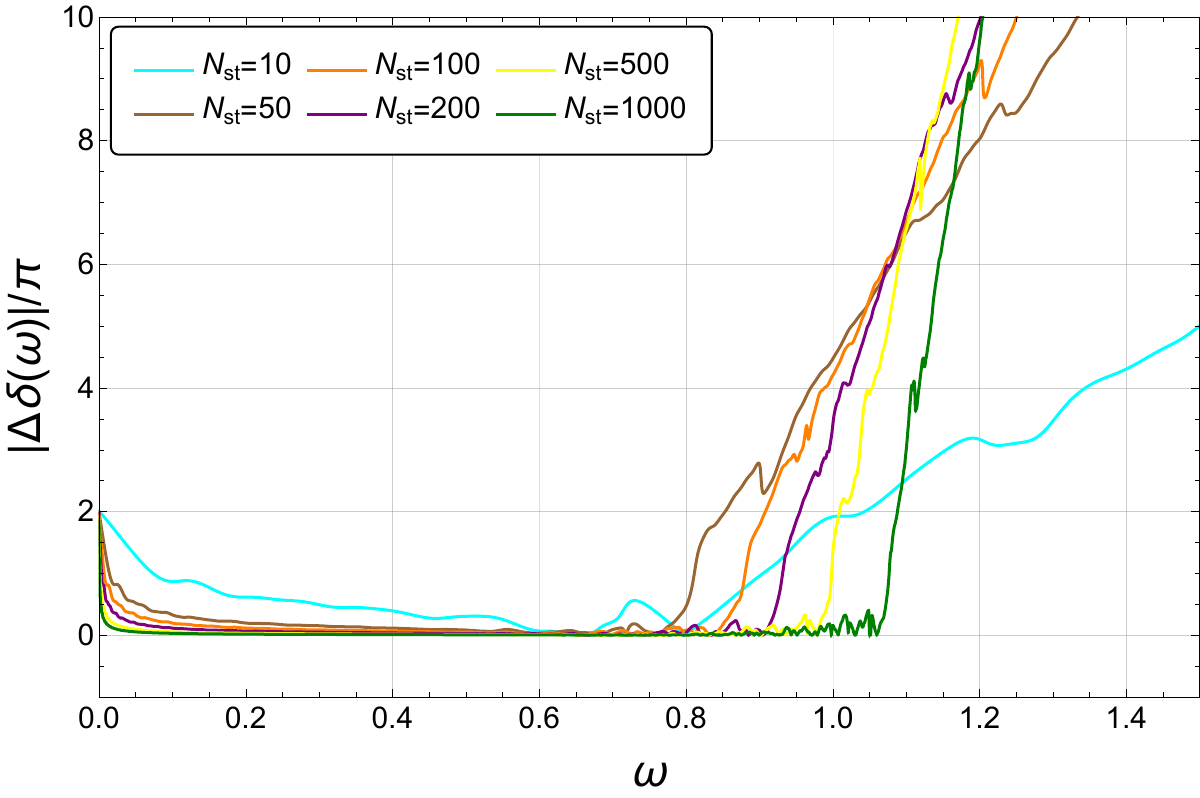}}}
	\caption{The magnitude of $H(\omega)$ and the phase $\delta(\omega)$ are shown in the left panels. The differences between $|H_P(\omega)|$ and $|H_S(\omega)|$, the differences between $\delta_P(\omega)$ and $\delta_S(\omega)$ are shown in the right panels. Different colors represent the piecewise step potential case with different $N_{\text{st}}$, where black color represents the Schwarzschild case. In the left bottom panel, the black dashed line corresponds to the result of subtracting $2\pi$ from the black solid line and is used to obtain the phase difference shown in the right bottom panel. Note that the phase unwrapping is imposed on the principal argument function $\Arg$ in the process of getting a continuous phase.} 
	\label{abs_H_omega_and_delta_omega}
\end{figure}

In the left panels of Fig. \ref{abs_H_omega_and_delta_omega}, the magnitude of $H(\omega)$ and the phase $\delta(\omega)$ are shown. In the right panels of Fig. \ref{abs_H_omega_and_delta_omega}, the differences between $|H_P(\omega)|$ and $|H_S(\omega)|$, the differences between $\delta_P(\omega)$ and $\delta_S(\omega)$ are shown, where $P$ refers to the piecewise step potential and $S$ refer to the Schwarzschild R-W potential. From the top panels of Fig. \ref{abs_H_omega_and_delta_omega}, it can be found that as $N_{\text{st}}$ increases, the difference in the magnitude of $H$ becomes progressively smaller, indicating that the magnitude of $H$ is stable. Since the reflection coefficient satisfy Eq. (\ref{reflection_amplitude_reflection_coefficient}), the reflection coefficient is also stable which is consistent with~\cite{Xie:2025jbr}, where piecewise parabolic approximation is used. For the case of a single bump external perturbation, the reflection coefficient also remains stable~\cite{Oshita:2024fzf,Wu:2024ldo,Rosato:2024arw,Ianniccari:2024ysv,Konoplya:2025ixm}. Therefore, for the reflection coefficients, they are stable in both piecewise approximation and small bump situations. From the bottom panels of Fig. \ref{abs_H_omega_and_delta_omega}, it can be seen that for a given $N_{\text{st}}$, as $\omega$ tends to zero, the phase $\delta_P(\omega)$ approaches $2\pi$. As $\omega$ tends to high-frequency, the difference between line $\delta_P(\omega)$ and line $\delta_S(\omega)-2\pi$ gradually increases. As $N_{\text{st}}$ increases, the phase of the piecewise step potential progressively converges with that of the Schwarzschild effective potential in the low-frequency regime, whereas the high-frequency components exhibit diametrically opposite behavior. In other words, the derivatives of $\delta_P(\omega)$ will be large at high-frequency regime. 

\begin{figure}[htbp]
    \centering
    \subfigure[]{\includegraphics[width=0.45\linewidth]{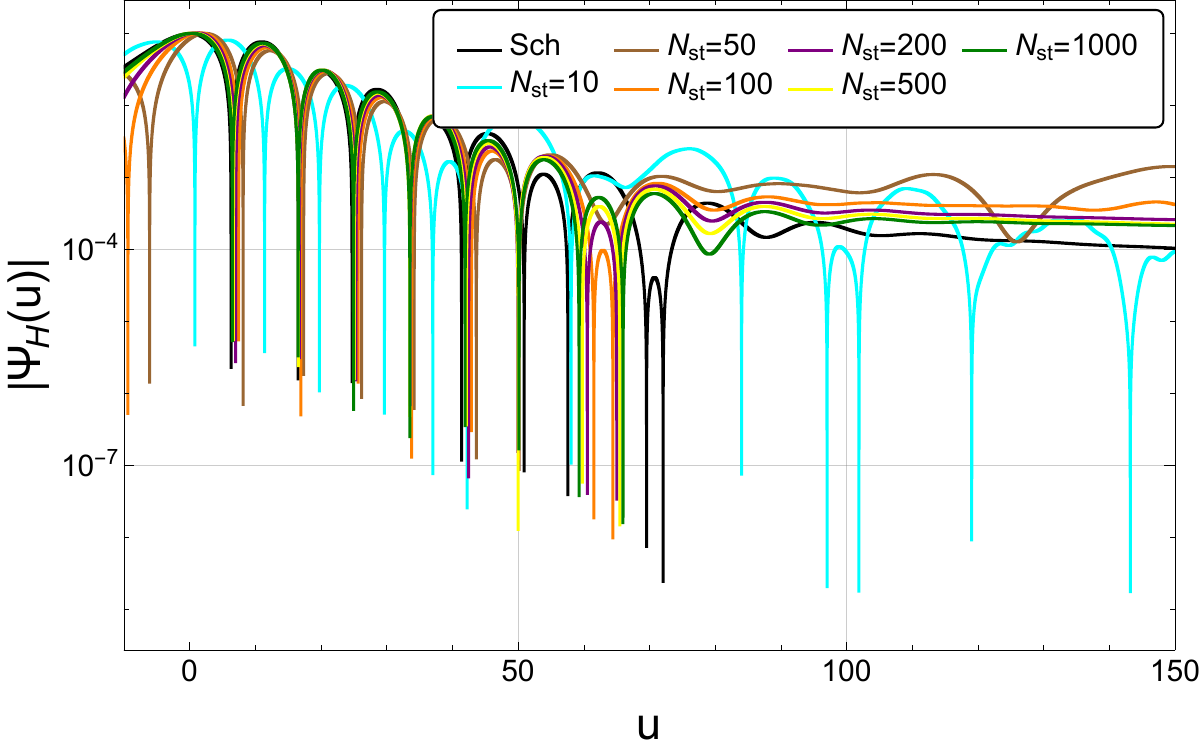}} \hfill
    \subfigure[]{\includegraphics[width=0.45\linewidth]{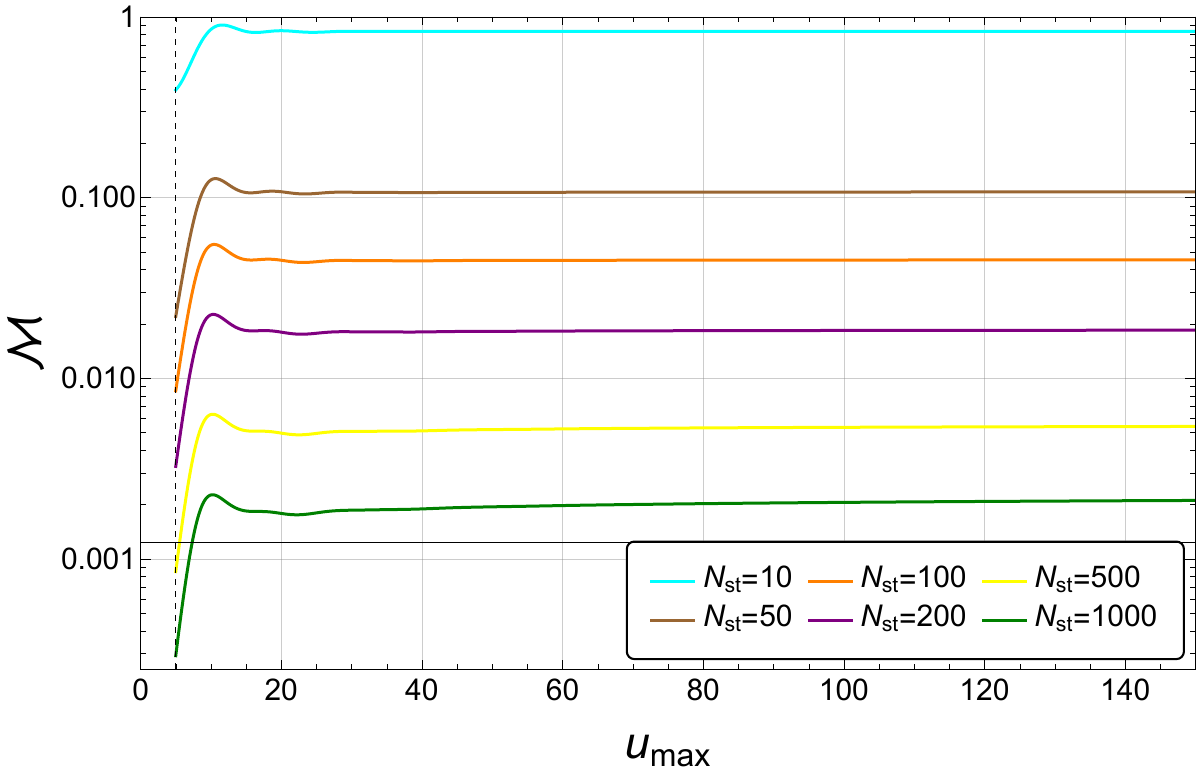}} 
    \caption{The left panel is the waveforms of $|\Psi_{\text{H},P}(u)|$ and $|\Psi_{\text{H},S}(u)|$. The right panel is the mismatch between $\Psi_{\text{H},P}(u)$ and $\Psi_{\text{H},S}(u)$ with the end time $u_{\text{max}}$ varying, where the dashed vertical line is $u_{\text{max}}=5$. The black color represents the result of Schwarzschild R-W potential, while other colors represent the results of piecewise step potential with different $N_{\text{st}}$.}
    \label{H_waveform_and_mismatch}
\end{figure}

Subsequently, one can make a comparison on the inverse transform of $H(\omega)$, i.e., the waveform $\Psi_{\text{H}}(u)$. These waveforms are shown in the left panel of Fig. \ref{H_waveform_and_mismatch}. As $N_{\text{st}}$ increases, the waveform of the piecewise step effective potential converges toward that of the Schwarzschild R-W potential, exhibiting a diminishing discrepancy. In the right panel of Fig. \ref{H_waveform_and_mismatch}, a quantitative description, mismatch, is also provided
\begin{eqnarray}\label{mismatch}
    \mathcal{M}[f(u),g(u)]=1-\frac{|\langle f(u),g(u)\rangle|}{\sqrt{\langle f(u),f(u)\rangle\langle g(u),g(u)\rangle}}\, ,
\end{eqnarray}
where the inner product is defined as
\begin{eqnarray}\label{inner_product}
    \langle f(u), g(u)\rangle=\int_{0}^{u_{\text{max}}}\mathrm{d}uf(u)\overline{g(u)}\, .
\end{eqnarray}
Here, $f$ refers to the waveform $\Psi_{\text{H},P}(u)$ and $g$ refers to the waveform $\Psi_{\text{H},S}(u)$. The integration is perform by the the composite trapezoidal formula with $\Delta u=0.0005$. Furthermore, we set the upper limit of the integral (\ref{inner_product}) to be variable, which means that the mismatch will be a function of $u_{\text{max}}$. From the right panel of Fig. \ref{H_waveform_and_mismatch}, increasing $N_{\text{st}}$ can reduce the mismatch. Therefore, under the initial condition of a far-distant Dirac delta source, the waveform remains stable.

We next concentrate on the waveform (\ref{fundamental_ringdown_F}), where the integrand is the transmission amplitude $F(\omega)$ whose reciprocal is $A_{\text{in}}(\omega)$. Actually, this waveform can be derived from a near-horizon Dirac delta source $I=2\mathrm{i}\omega\delta(x-x_s)$ with $x_s\ll x_m$ [see Eq. (\ref{waveform_sourced_delta_event})]. Similar to $H(\omega)$, we also conduct a three-way comparison, specifically in magnitude, phase, and corresponding waveform [see Eq. (\ref{fundamental_ringdown_F})]. However, when it comes to the waveform $\Psi_{\text{F}}(u)$, particular care must be taken in performing the integration. Since the high frequency behavior of $F(\omega)$ is $|F(\omega)|\to1$, the function $\lvert F(\omega)\rvert$ is not a square-integrable function, in contrast to the case of $H(\omega)$. To circumvent this issue, we assume the function $F(\omega)$ tends to a constant $e^{i d_0}$ with a real $d_0$ as $\omega\to+\infty$, and construct a square-integrable function as
\begin{eqnarray}
    {}^0F(\omega)\equiv F(\omega)-\mathrm{e}^{\mathrm{i} d_0}\quad \text{for}\quad  \omega\geq0\, ,
\end{eqnarray}
which tends to $0$ as $\omega\to+\infty$.
The $\omega < 0$ component of the function ${}^0F(\omega)$ is determined by requiring that ${}^0F(\omega)$ possesses the same symmetry as $F(\omega)$, namely $F(-\omega) = \overline{F(\omega)}$. Actually, one should aware that the waveform $\Psi_{\text{F}}(u)$ contains a delta function $\delta(u)$~\cite{Mark:2017dnq}, and the waveform ${}^0\Psi_{\text{F}}(u)$ obtained by the Fourier transformation of ${}^0F(\omega)$ excludes the delta function $\delta(u)$,
\begin{eqnarray}
    \Psi_{\text{F}}(u)&=&\frac{1}{2\pi}\int_{-\infty}^{+\infty}\mathrm{d}\omega \Big[{}^0F(\omega)+\Theta(\omega)\mathrm{e}^{\mathrm{i} d_0}+\Theta(-\omega)\mathrm{e}^{-\mathrm{i} d_0}\Big] \mathrm{e}^{-\mathrm{i}\omega u}\nonumber\\
    &=&{}^0\Psi_{\text{F}}(u)+\cos (d_0)\delta(u)+\frac{\sin(d_0)}{\pi u}\, ,
\end{eqnarray}
where
\begin{eqnarray}\label{Psi0}  
    {}^0\Psi_{\text{F}}(u)&\equiv& \frac{1}{\pi}\int_{0}^{+\infty}\mathrm{d}\omega \Re\Big[{}^0F(\omega)\mathrm{e}^{-\mathrm{i}\omega u}\Big]\approx\, \frac{1}{\pi}\int_{0}^{\omega_m}\mathrm{d}\omega \Re\Big[{}^0F(\omega)\mathrm{e}^{-\mathrm{i}\omega u}\Big]\, .
\end{eqnarray}
In the above expression, $\omega_m$ denotes a numerical cutoff, and numerically we take $d_0=\eta(\omega_m)$. Based on the above, we are more inclined to discuss the stability of waveform ${}^0\Psi_{\text{F}}(u)$.

\begin{figure}[htbp]
    \centering
    \subfigure[]
	{\includegraphics[width=0.45\textwidth]{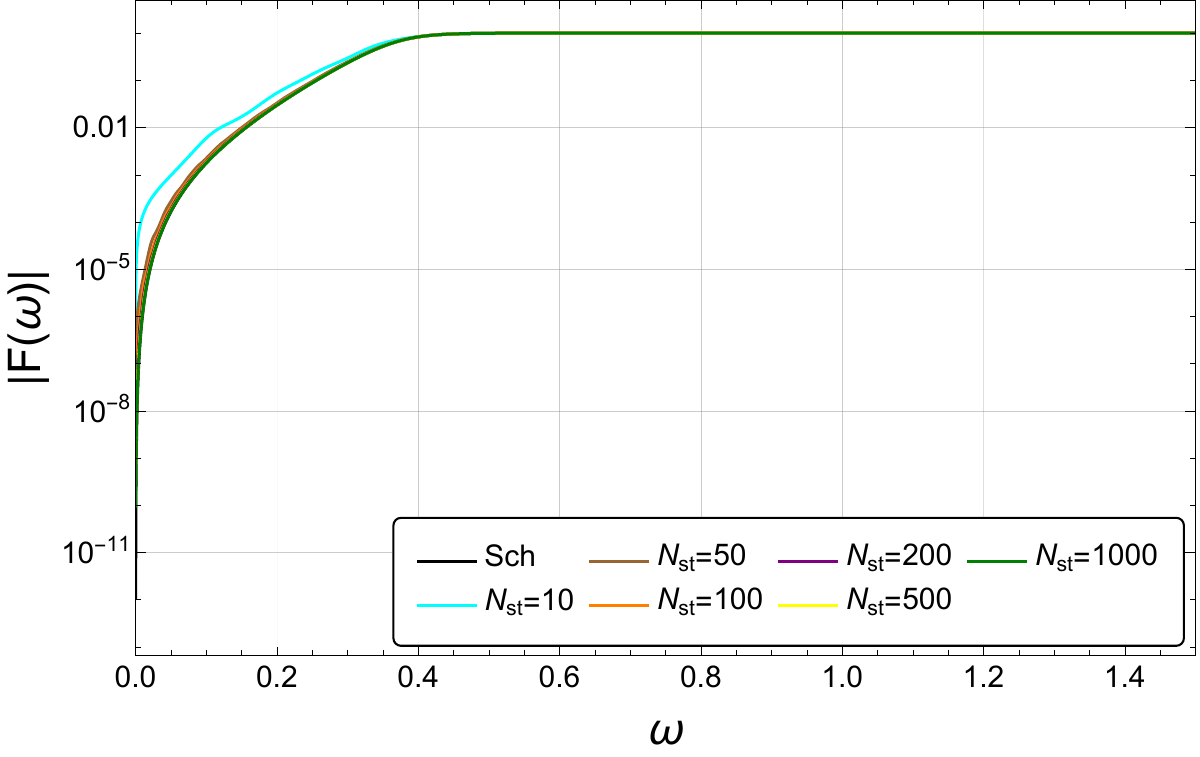}\label{Abs_F_omega}} \hfill
    \subfigure[]
	{\includegraphics[width=0.45\textwidth]{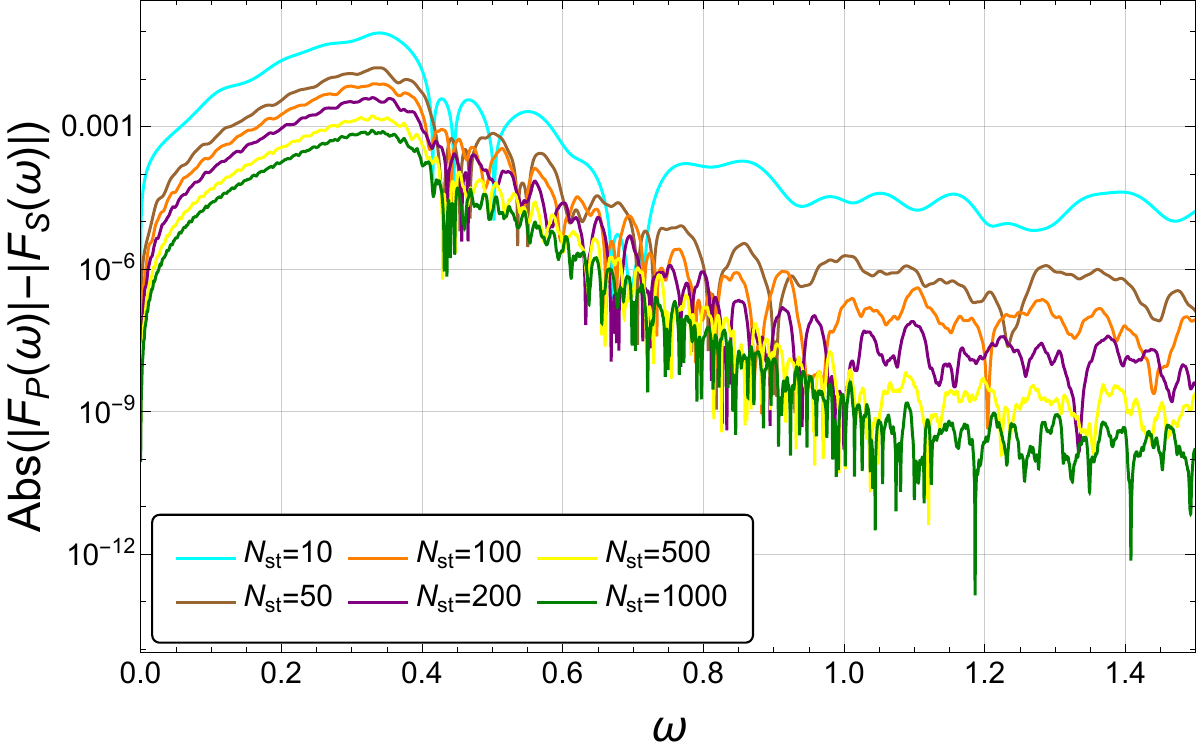}}
    \subfigure[]
	{\includegraphics[width=0.45\textwidth]{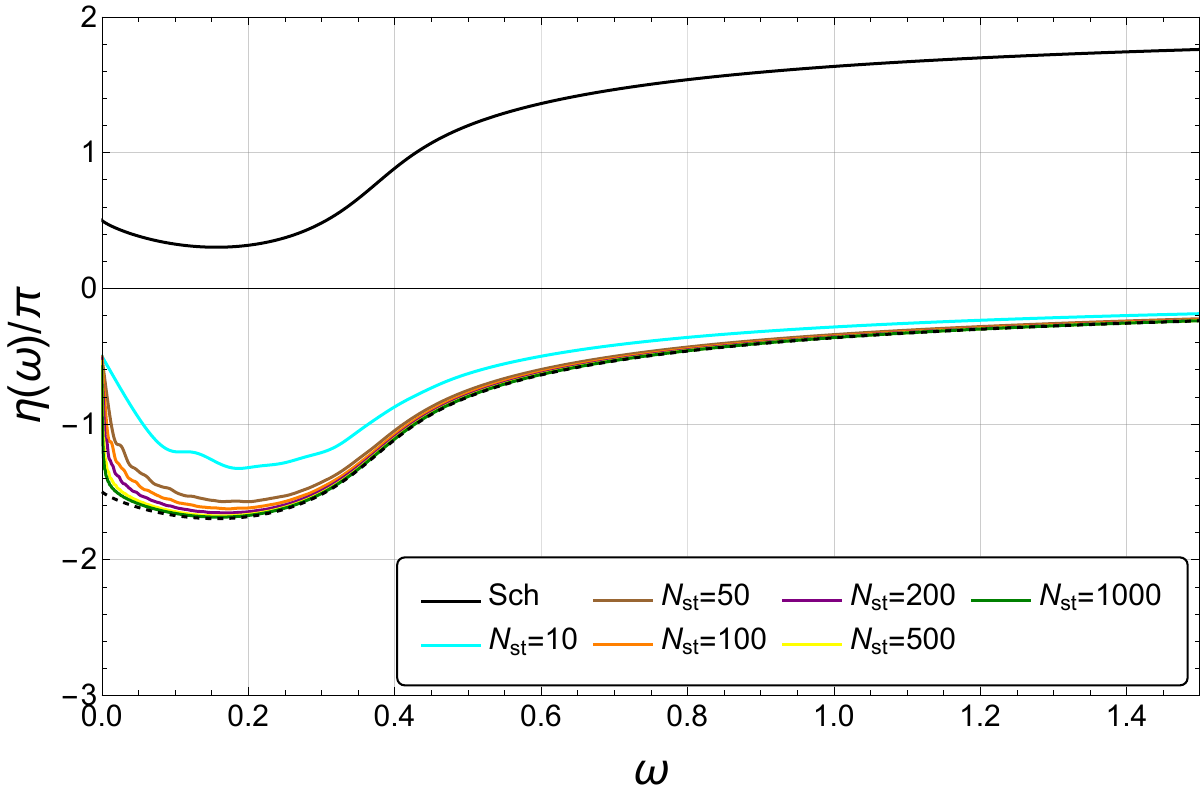}} \hfill
    \subfigure[]
	{\includegraphics[width=0.45\textwidth]{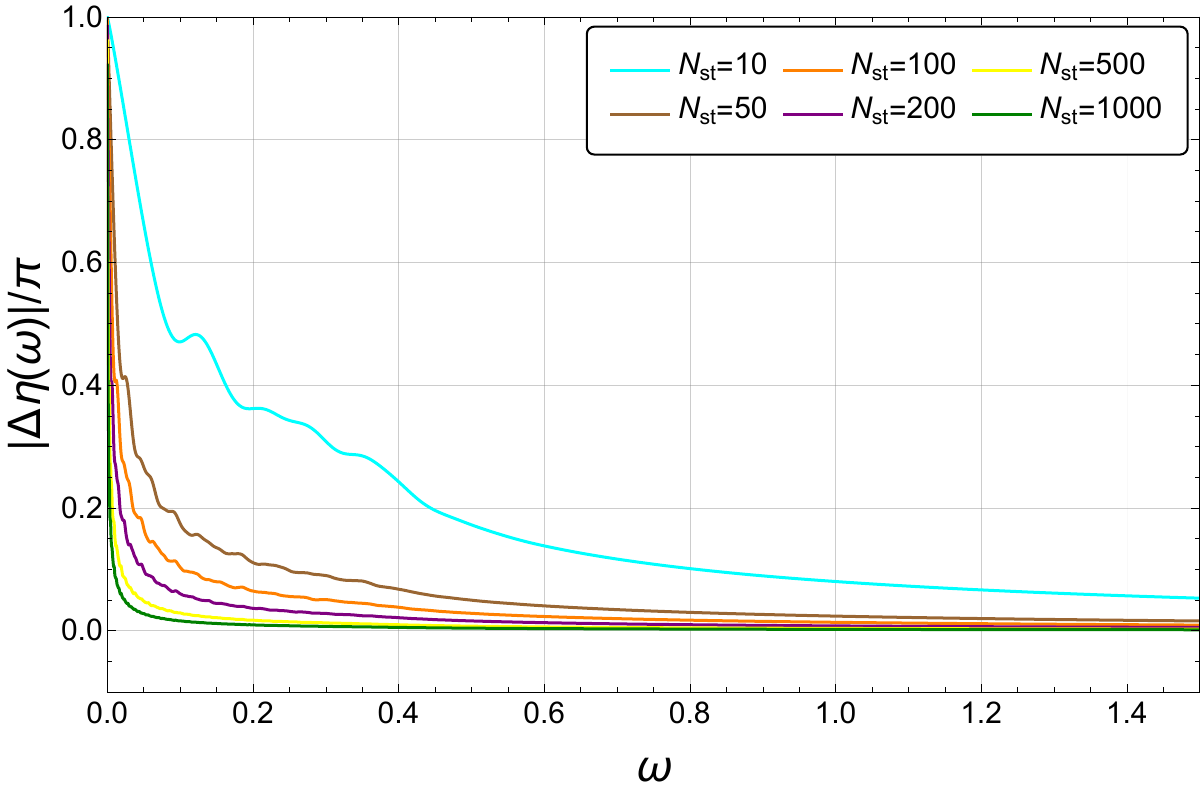}}
    \caption{The magnitude of $F(\omega)$ and the phase $\eta(\omega)$ are shown in the left panels. The differences between $|F_P(\omega)|$ and $|F_S(\omega)|$, the differences between $\eta_P(\omega)$ and $\eta_S(\omega)$ are shown in the right panels. Different colors represent the piecewise step potential case with different $N_{\text{st}}$, where black color represents the Schwarzschild case. In the left bottom panel, the black dashed line corresponds to the result of subtracting $2\pi$ from the Schwarzschild phase.}
    \label{abs_F_omega_and_eta_omega}
\end{figure}

In the left panels of Fig. \ref{abs_F_omega_and_eta_omega}, the magnitude of $F(\omega)$, the phase $\eta(\omega)$ are shown. Note that our results of the phase $\eta(\omega)$ for the Schwarzschild case is in agreement with~\cite{Kyutoku:2022gbr} (see left top panel of Fig. 7 therein). In the right panels of Fig. \ref{abs_F_omega_and_eta_omega}, the differences between $|F_P(\omega)|$ and $|F_S(\omega)|$ with the differences $\Delta\eta(\omega)=\eta_P(\omega)-\eta_P(\omega)+2\pi$ are shown. The upper two panels of Fig. \ref{abs_F_omega_and_eta_omega} demonstrate that the discrepancy in $F$'s magnitude diminishes progressively with increasing $N_{\text{st}}$, suggesting a stable magnitude characteristics for $F$. Given the relationship $\Gamma(\omega)=|F(\omega)|^2$, the graybody factor, $\Gamma(\omega)$, consequently exhibits stable characteristic~\cite{Xie:2025jbr}. With increasing $N_{\text{st}}$, the phase of the piecewise step potential asymptotically approaches that of the Schwarzschild effective potential within the low-frequency domain, while analogous convergence behavior is observed in the high-frequency regime.

\begin{figure}[htbp]
    \centering
    \subfigure[]{\includegraphics[width=0.45\linewidth]{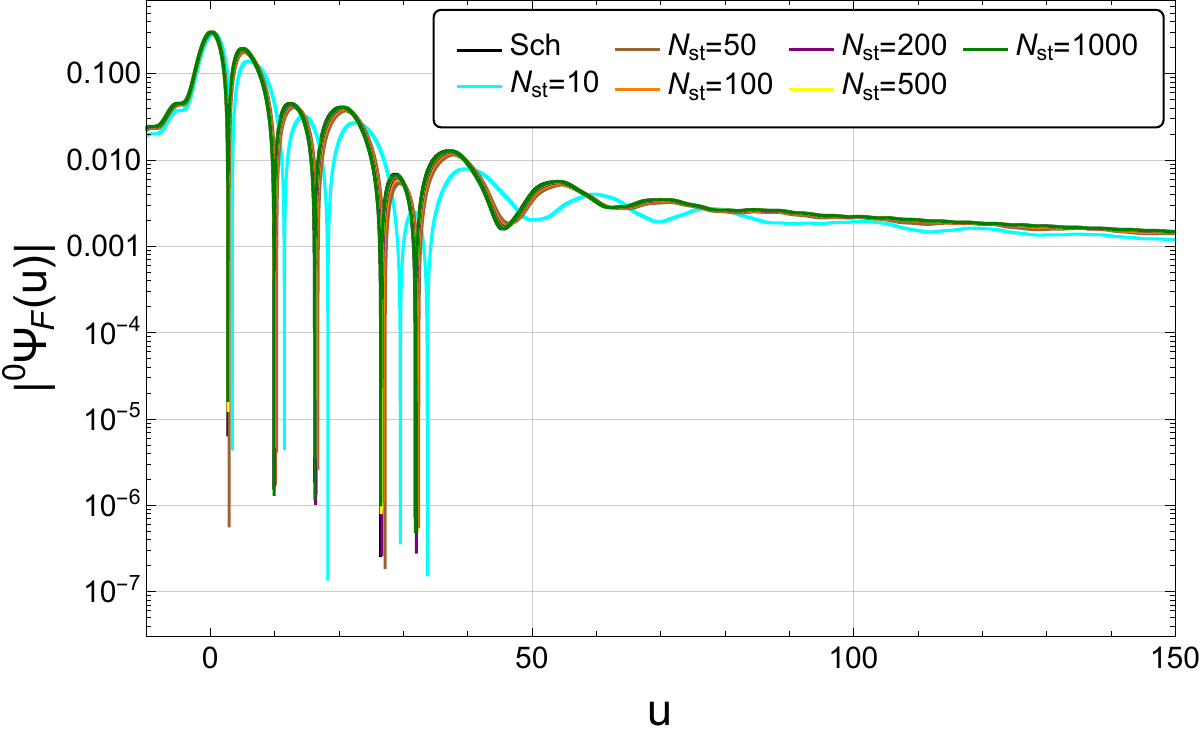}} \hfill
    \subfigure[]{\includegraphics[width=0.45\linewidth]{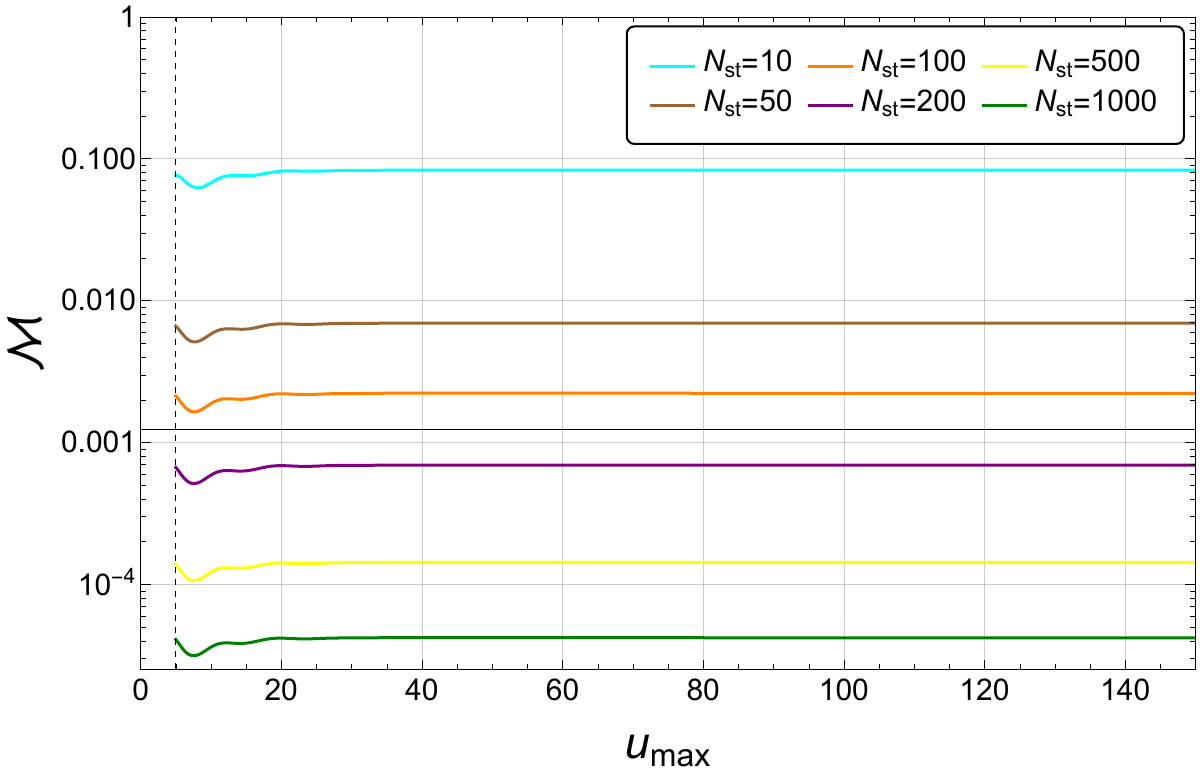}} 
    \caption{The left panel is the waveforms of $|{}^0\Psi_{\text{F},P}(u)|$ and $|{}^0\Psi_{\text{F},S}(u)|$. The right panel is the mismatch between ${}^0\Psi_{\text{F},P}(u)$ and ${}^0\Psi_{\text{F},S}(u)$ with the end time $u_{\text{max}}$ varying, where the dashed vertical line is $u_{\text{max}}=5$. The black color represents the result of Schwarzschild R-W potential, while other colors represent the results of piecewise step potential with different $N_{\text{st}}$.}
    \label{F_waveform_and_mismatch}
\end{figure}

Subsequent comparative analysis focuses on the waveform ${}^0\Psi_{\text{F}}(u)$. As illustrated in the left panel of Fig. \ref{F_waveform_and_mismatch}, increasing $N_{\text{st}}$ drives the piecewise step potential's waveform toward progressive convergence with the Schwarzschild R-W potential, exhibiting a monotonic reduction in waveform deviation. The right panel of Fig. \ref{F_waveform_and_mismatch} quantitatively characterizes this convergence through mismatch (\ref{mismatch}). Here, $f$ is the waveform ${}^0\Psi_{\text{F},P}(u)$ and $g$ is the waveform ${}^0\Psi_{\text{F},S}(u)$. Therefore, under the initial condition of a Dirac delta source with the source positioned such that $x_s\ll x_m$, the waveform remains stable. From right panel of Fig. \ref{H_waveform_and_mismatch} and right panel of Fig. \ref{F_waveform_and_mismatch}, quantitative comparison further indicates that the Dirac delta source near the event horizon has significantly better waveform stability compared to the Dirac delta source near infinity.

\begin{figure}[htbp]
    \centering
    \subfigure[]{\includegraphics[width=0.45\linewidth]{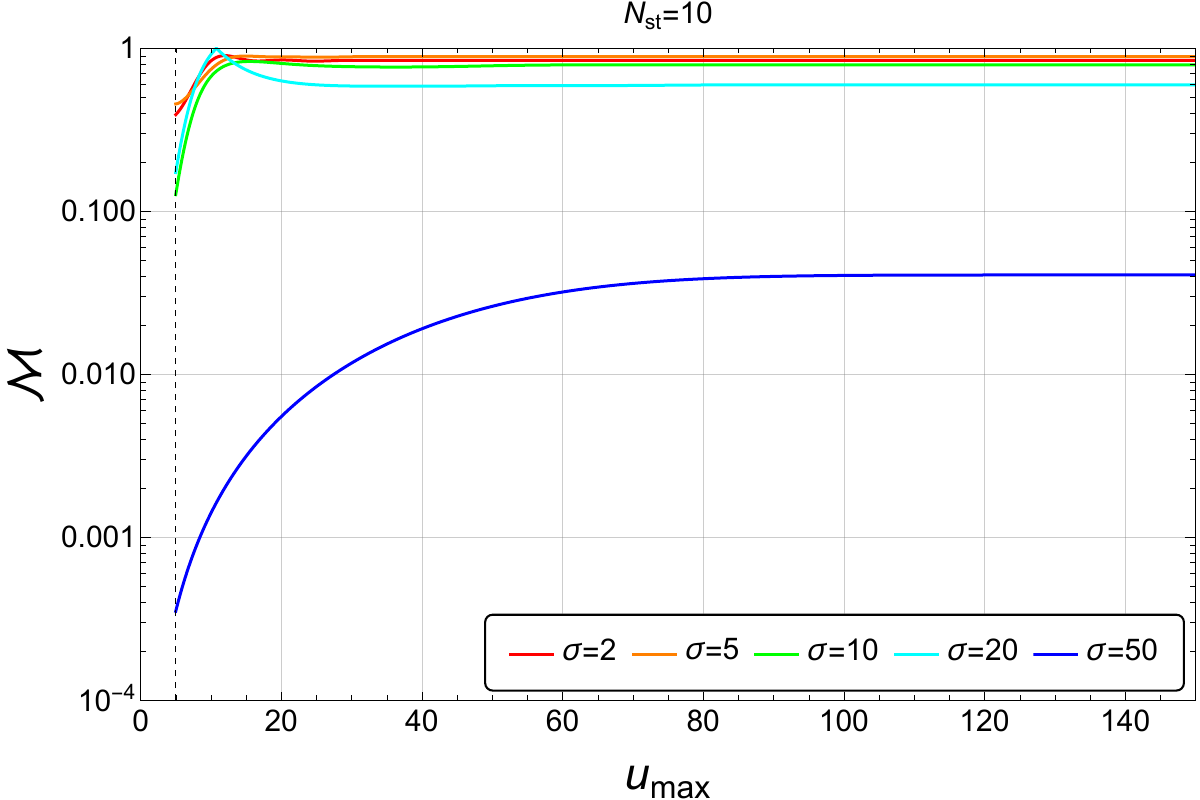}}  \hfill
    \subfigure[]{\includegraphics[width=0.45\linewidth]{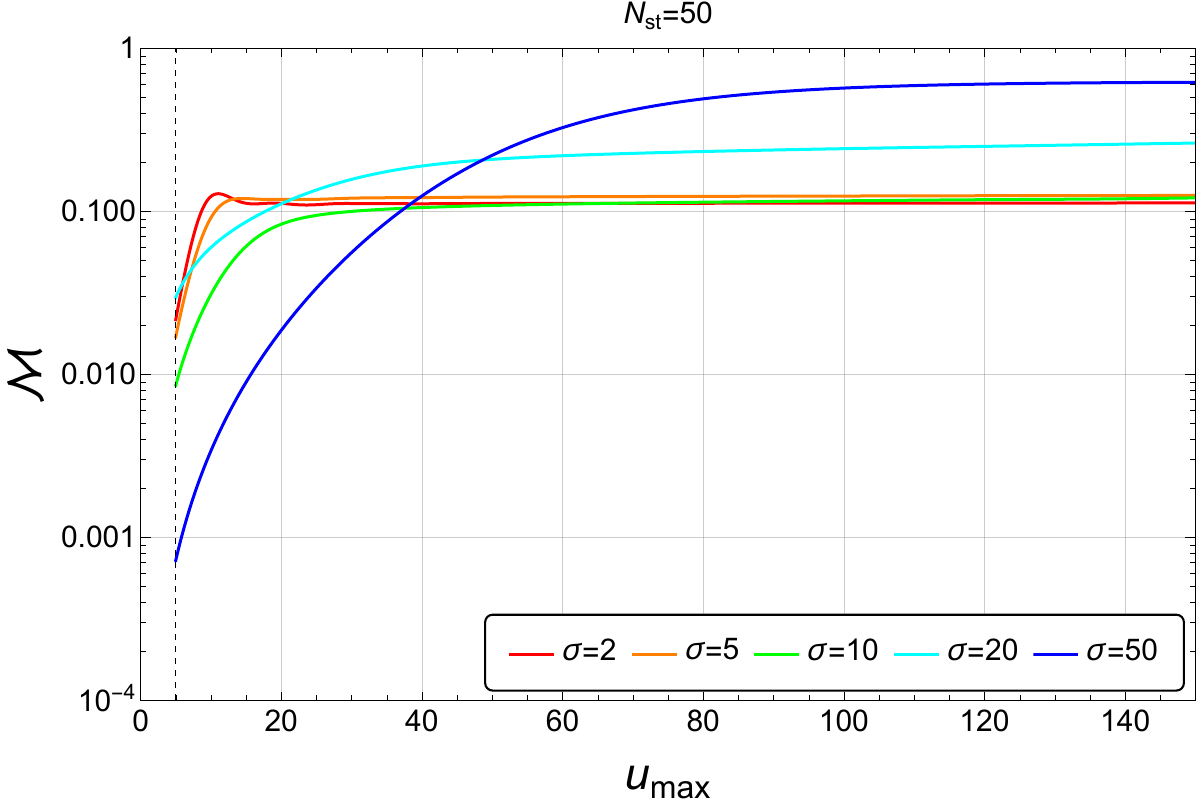}}  \\
    \subfigure[]{\includegraphics[width=0.45\linewidth]{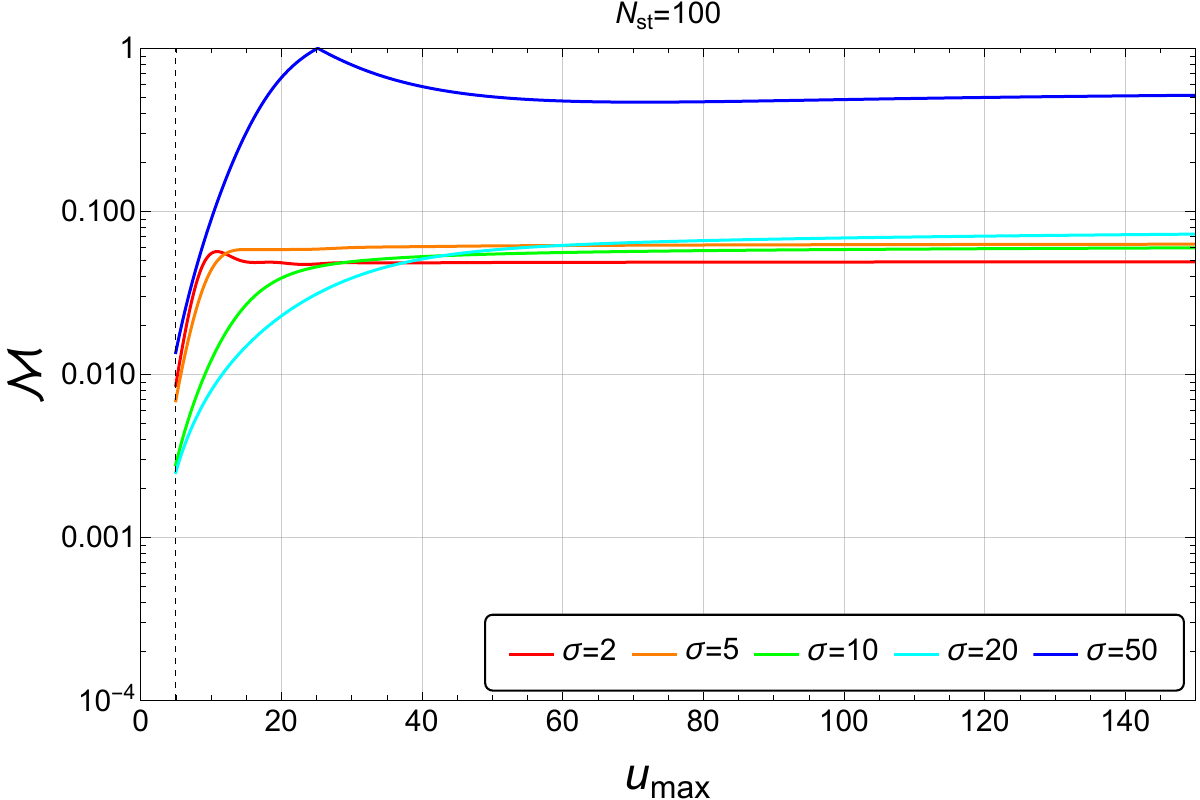}}  \hfill
    \subfigure[]{\includegraphics[width=0.45\linewidth]{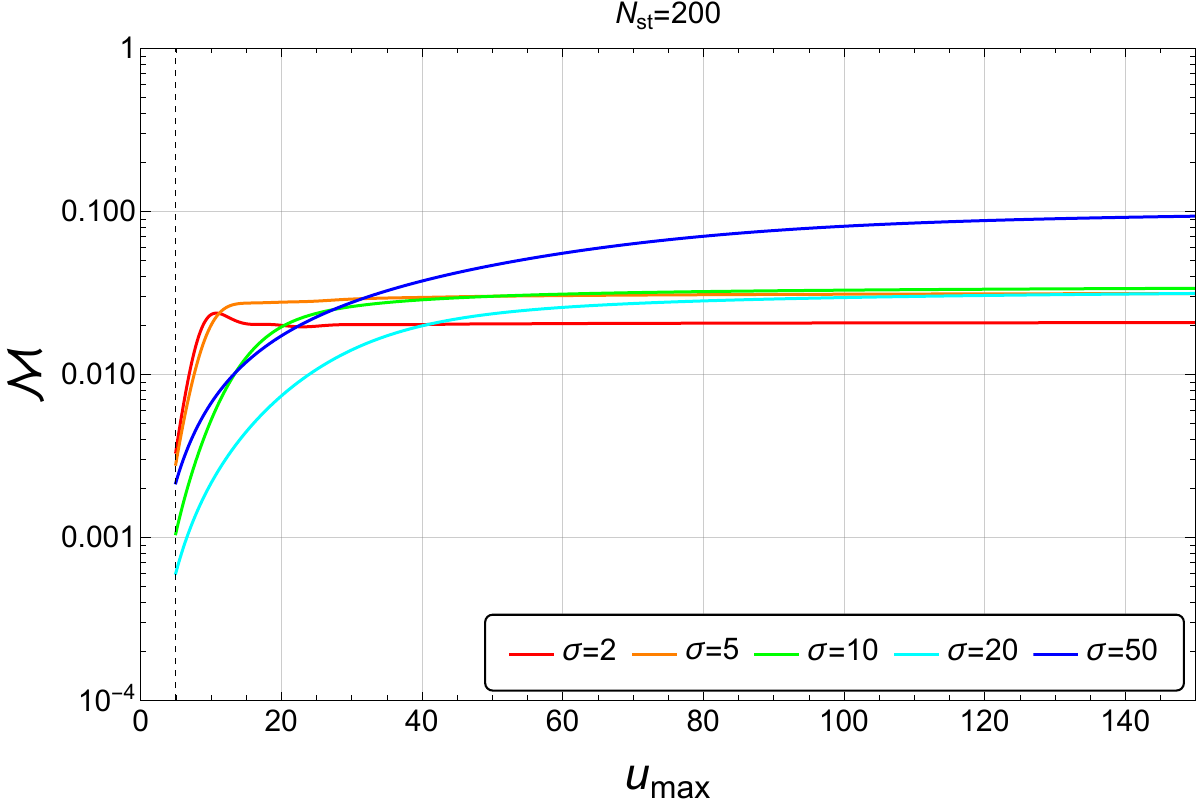}}  \\
    \subfigure[]{\includegraphics[width=0.45\linewidth]{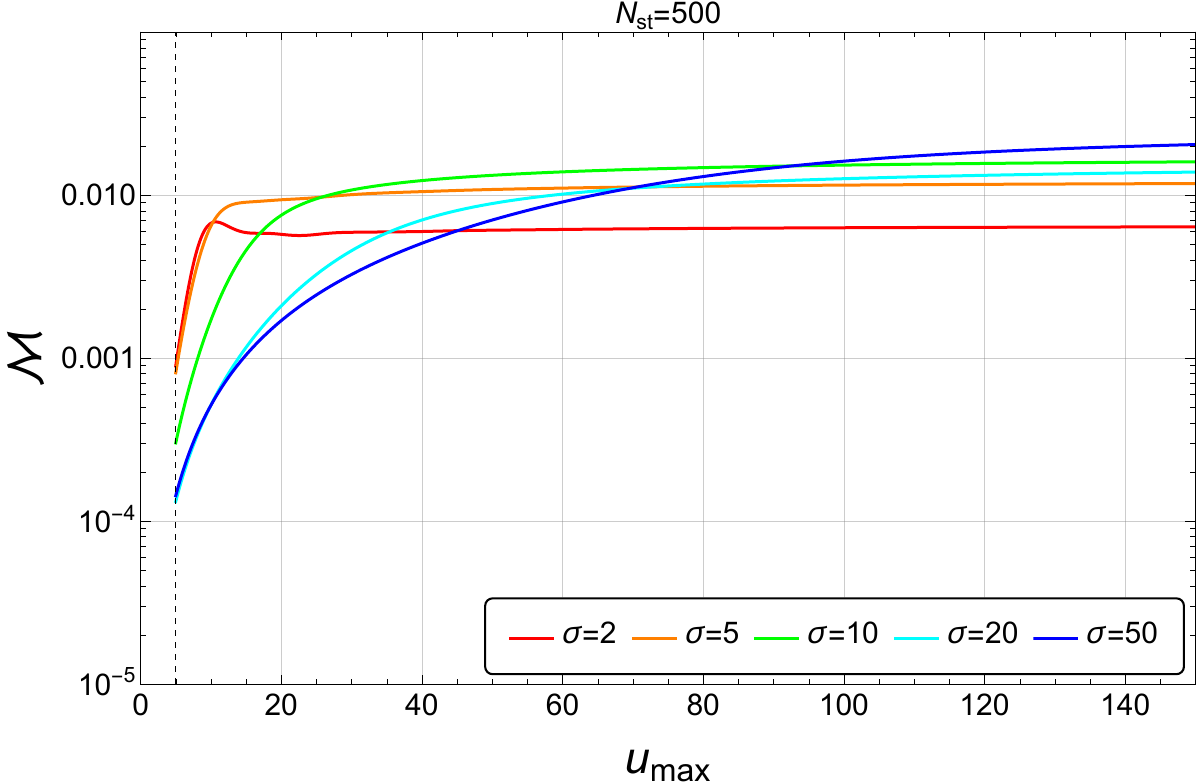}}  \hfill
    \subfigure[]{\includegraphics[width=0.45\linewidth]{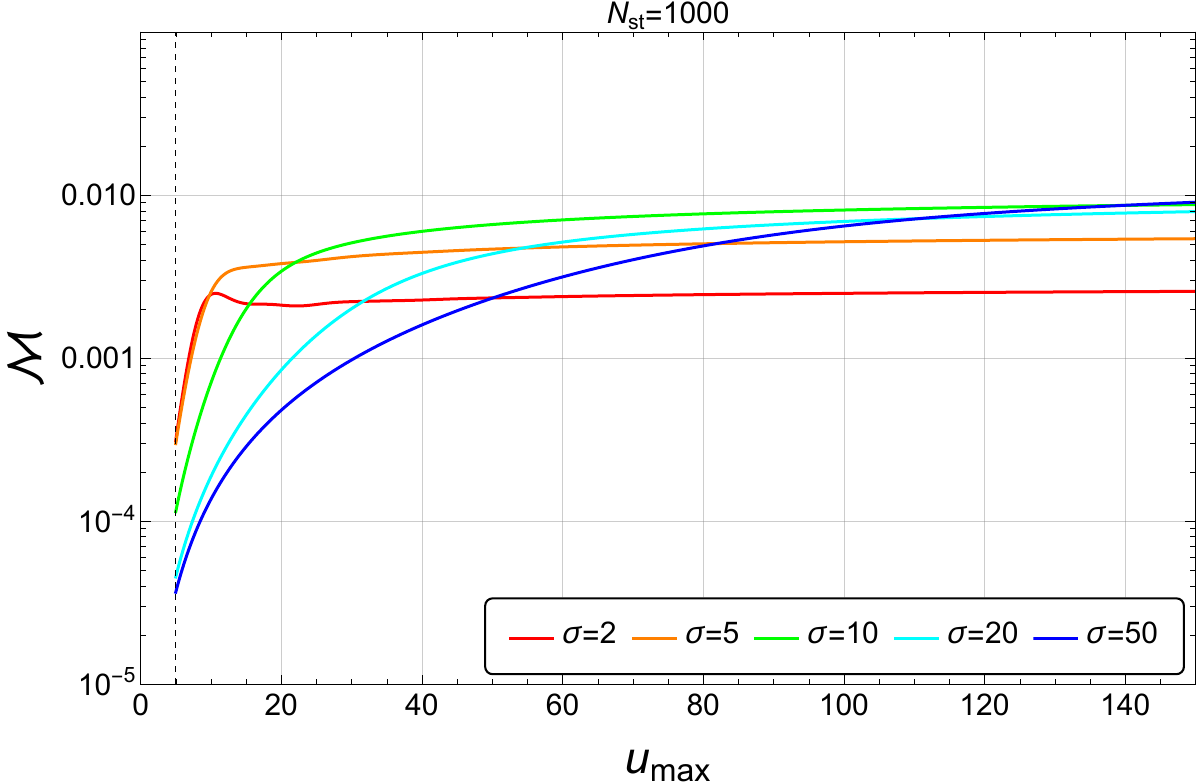}\label{H_TG_mismatch_1000}}  \\
    \caption{The mismatches between the waveforms $\Psi_{\text{H},P}^{\text{G}}(u)$ and $\Psi_{\text{H},S}^{\text{G}}(u)$ are shown. For each panels, its parameter $N_{\text{st}}$ is fixed, where we choose $N_{\text{st}}=10$, $50$, $100$, $200$, $500$ and $1000$ as our outputs. In each panel, different color lines correspond to different widths $\sigma$ of the initial Gauss bumps (\ref{initial_Gauss_bump}).}
    \label{H_TG_mismatch}
\end{figure}

\begin{figure}[htbp]
    \centering
    \subfigure[]{\includegraphics[width=0.45\linewidth]{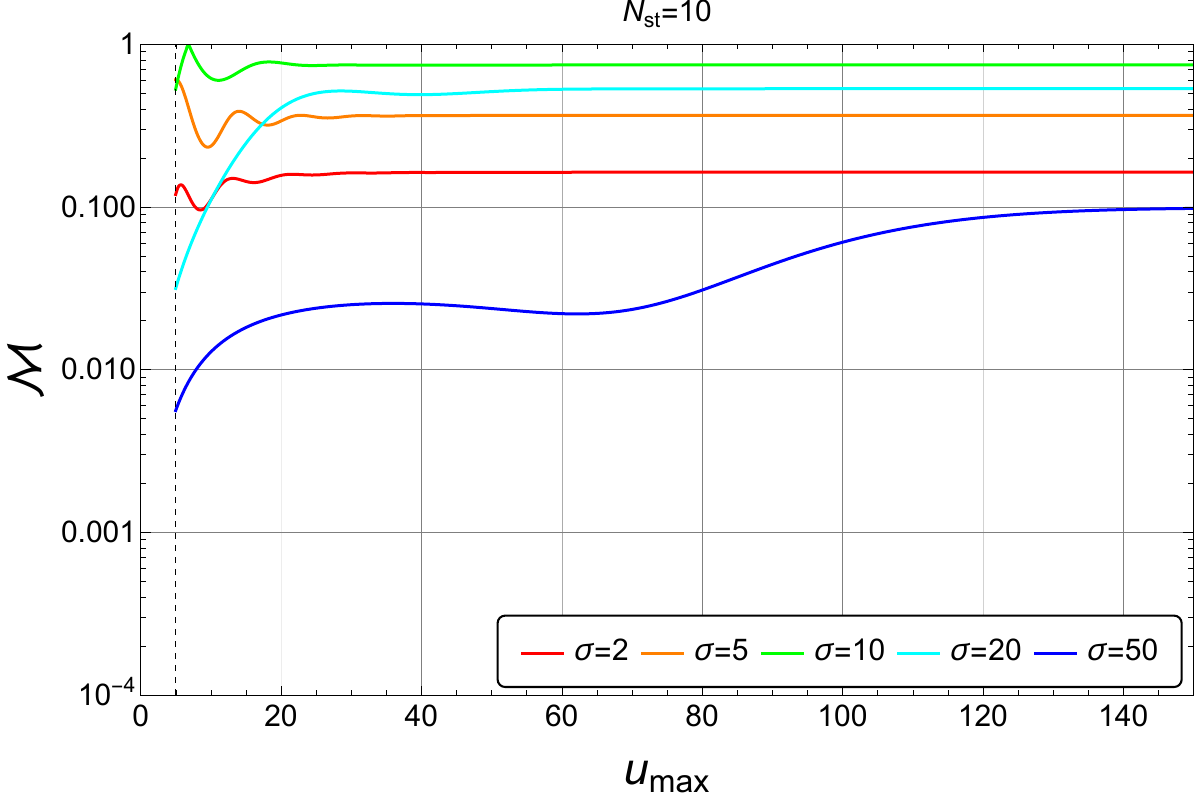}}  \hfill
    \subfigure[]{\includegraphics[width=0.45\linewidth]{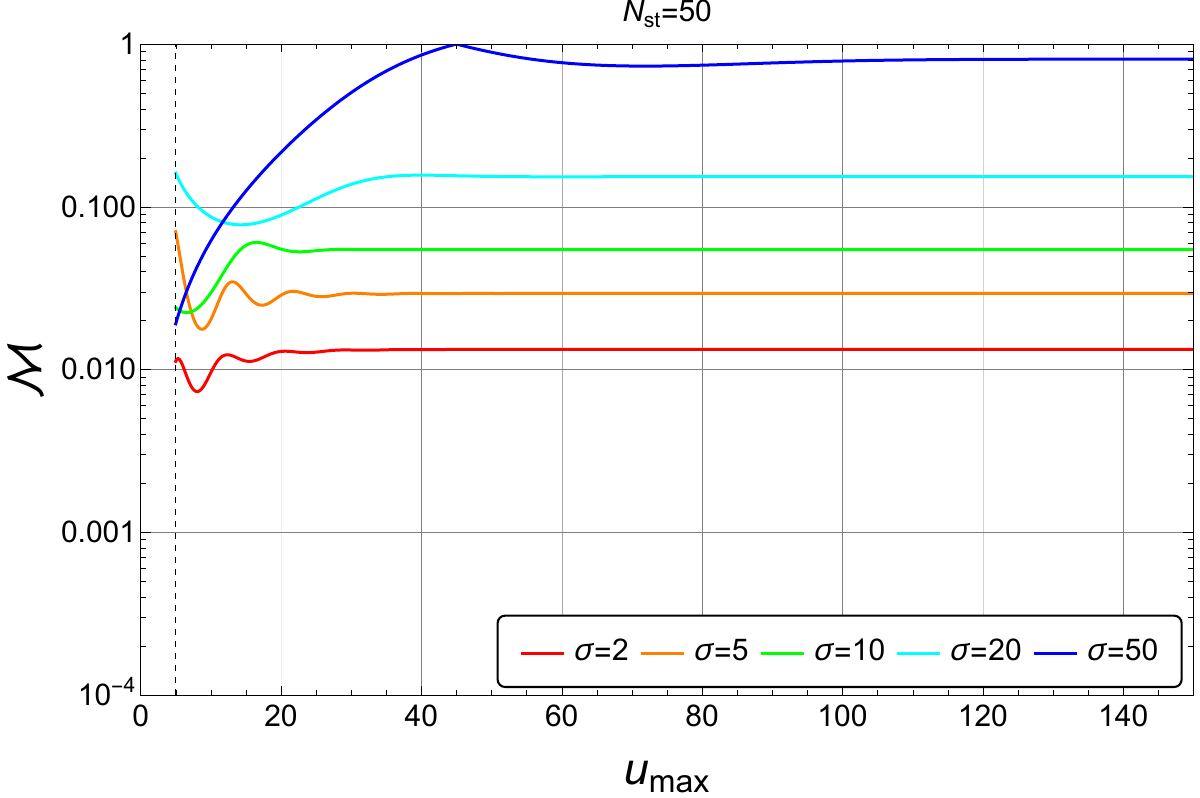}}  \\
    \subfigure[]{\includegraphics[width=0.45\linewidth]{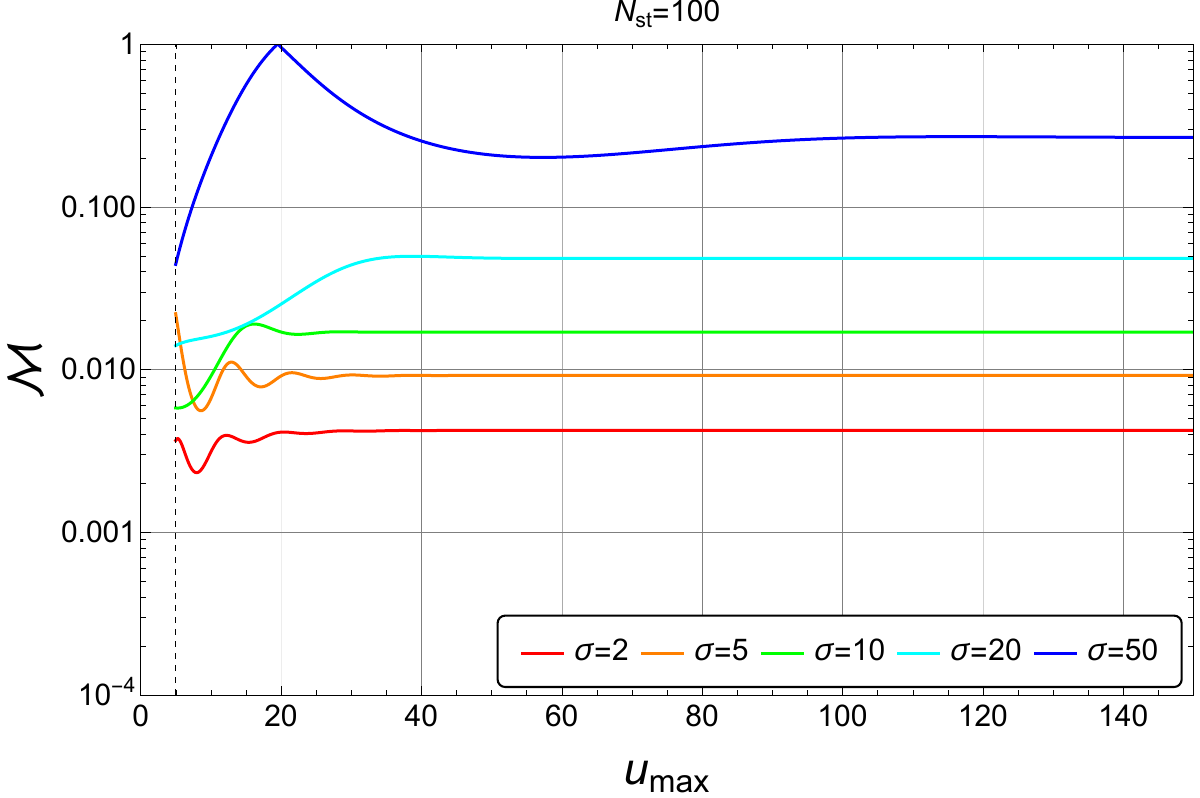}}  \hfill
    \subfigure[]{\includegraphics[width=0.45\linewidth]{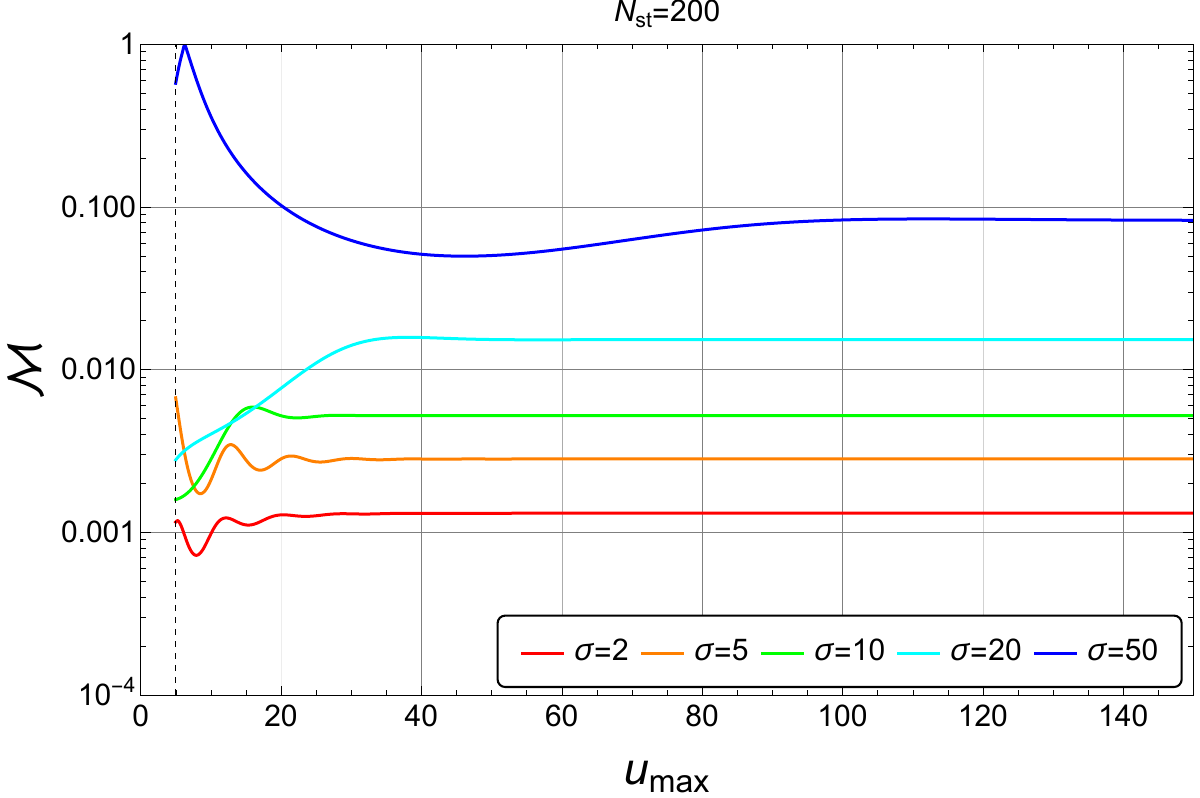}}  \\
    \subfigure[]{\includegraphics[width=0.45\linewidth]{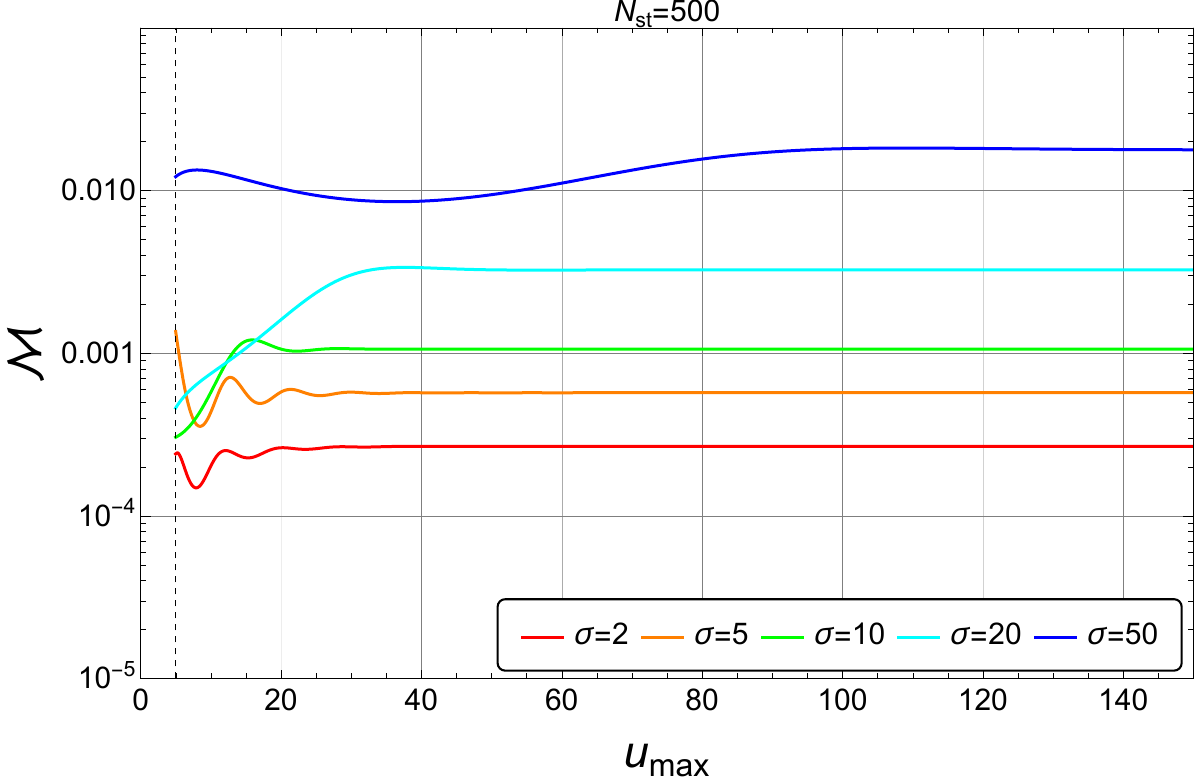}}  \hfill
    \subfigure[]{\includegraphics[width=0.45\linewidth]{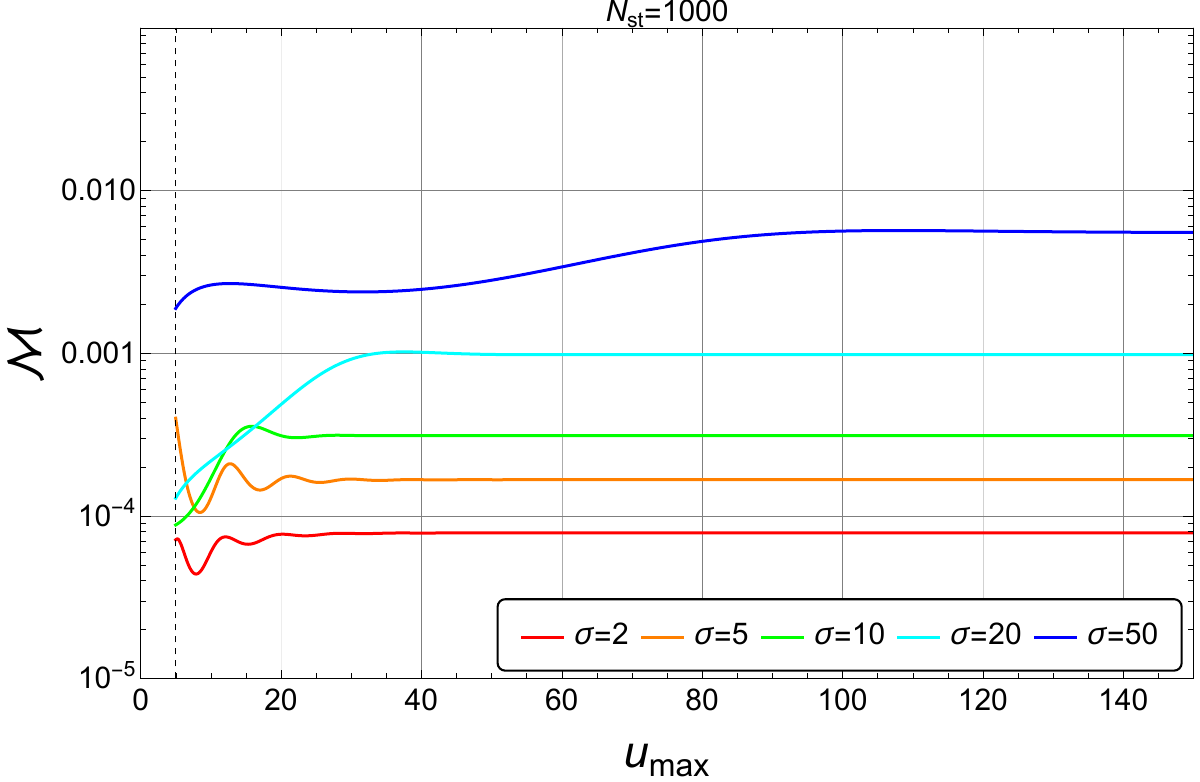}\label{F_TG_mismatch_1000}}  \\
    \caption{The mismatches between the waveforms $\Psi_{\text{F},P}^{\text{G}}(u)$ and $\Psi_{\text{F},S}^{\text{G}}(u)$ are shown. In each panel, the parameter $N_{\text{st}}$ is fixed, where we choose $N_{\text{st}}=10$, $50$, $100$, $200$, $500$ and $1000$ as our outputs. Within each panel, curves of different colors correspond to different widths $\sigma$ of the initial Gauss bumps (\ref{initial_Gauss_bump}).}
    \label{F_TG_mismatch}
\end{figure}

Considering a more realistic situation, we now examine the influence of Gauss bump sources $(\sigma>0)$ on waveform stability. This represents a more general case compared to our prior investigation into the impact of Dirac delta sources. Notably, when the width of the Gaussian source approaches zero, the scenario converges precisely to the Dirac delta source condition (cf. Appendix \ref{sec: waveforms_sourced_by_Gauss_bump}). More specifically, we study the following waveforms
\begin{eqnarray}\label{Psi_G_H}
    \Psi_{\text{H}}^{\text{G}}(u)=\frac{1}{2\pi}\int_{-\infty}^{+\infty}\mathrm{d}\omega H(\omega) T_{\text{G}}(\omega) \mathrm{e}^{-\mathrm{i}\omega u}\, ,
\end{eqnarray}
and
\begin{eqnarray}\label{Psi_G_F}
    \Psi_{\text{F}}^{\text{G}}(u)=\frac{1}{2\pi}\int_{-\infty}^{+\infty}\mathrm{d}\omega F(\omega) T_{\text{G}}(\omega) \mathrm{e}^{-\mathrm{i}\omega u}\, .
\end{eqnarray}
The function $T_{\text{G}}(\omega)$ is a Gauss bump with center at zero-frequency and width $1/\sigma$ [see Eq. (\ref{T_G_omega})]. The function $T_{\text{G}}(\omega)$ can be viewed as a low-pass filter, with a larger $\sigma$ corresponding to a lower cutoff frequency and greater attenuation of high-frequency components. It is worth mentioning that due to the existence of $T_{\text{G}}(\omega)$, the integrand $F(\omega)T_{\text{G}}(\omega)$ in Eq. (\ref{Psi_G_F}) is square integrable, so its inverse transformation can be well discussed. Since the inverse Fourier transformation is truncated at $\omega_m = 1.5 \gg \sqrt{V_m}$~\cite{Rosato:2025byu}, the parameter $\sigma$ in $T_{\text{G}}(\omega)$ is required to satisfy $\sigma \gtrsim 2$ under the $3$-$\sigma$ principle to guarantee numerical accuracy. For the waveform $\Psi_{\text{H}}^{\text{G}}(u)$, the mismatches are shown in Fig. \ref{H_TG_mismatch}, and for the waveform $\Psi_{\text{F}}^{\text{G}}(u)$ the mismatches are shown in Fig. \ref{F_TG_mismatch}. In each figures, there are six subfigures. In each subfigure, parameter $N_{\text{st}}$ is fixed and different colors represent the mismatch with different widths $\sigma$, namely $\sigma=2$, $5$, $10$, $20$, and $50$.

In Fig. \ref{H_TG_mismatch}, on the one hand, if we focus on comparing lines of the same color across different panels just fixing $\sigma$, it can still be concluded that the mismatch function decreases as $N_{\text{st}}$ increases. This trend is similar to that observed in the case of a delta source (see Fig. \ref{H_waveform_and_mismatch}). The distinction lies in the fact that for the delta source, the discrete tiers of $\mathcal{M}$ are clearly distinguishable, whereas for the Gauss bump source, it only shows a general decreasing trend. Especially for the case of $\sigma=50$, the variation of $\mathcal{M}$ is the most complex. On the other hand, for a given parameter $N_{\text{st}}$, the mismatch function exhibits a roughly monotonic increase with growing $u_{\text{max}}$, especially for $\sigma= 5$, $10$, $20$, and $50$. As for $\sigma= 2$, it is found that the mismatch function is not significantly different from the corresponding delta source case $(\sigma=0)$. 

A comparison of identically colored lines across the panels in Fig. \ref{F_TG_mismatch} shows that, for a given width $\sigma$, the mismatch between the waveforms $\Psi^{\text{G}}_{\text{F},P}(u)$ and $\Psi^{\text{G}}_{\text{F},S}(u)$ decreases as $N_{\text{st}}$ increases. This trend is similar to that observed in the case of a delta source (see Fig. \ref{F_waveform_and_mismatch}). However, the result of $\sigma=50$ is an exception. When $\sigma= 2$, no significant difference is observed between the mismatch function for the Gauss bump source and the corresponding delta source mismatch function, which is same as Fig. \ref{H_TG_mismatch}. A noticeable distinction between Fig. \ref{F_TG_mismatch} and Fig. \ref{H_TG_mismatch} is that the variety range of the mismatch function associated with $u_{\text{max}}$ in Fig. \ref{F_TG_mismatch} is not as large as in Fig. \ref{H_TG_mismatch}.

Given that the implications of minor alterations in the effective potential are particularly capable of arousing significant research interest, as exemplified by the slight modifications induced by certain quantum effects, the results provided by $N_{\text{st}}=1000$ are consequently more valuable. From Fig. \ref{H_TG_mismatch_1000}, for the reflection ringdown waveform $\Psi_{\text{H}}^{\text{G}}(u)$, when $u_{\text{max}}$ is small, the mismatch function of small $\sigma$ is larger, whereas when $u_{\text{max}}$ is large, the mismatch function of large $\sigma$ is larger. From Fig. \ref{F_TG_mismatch_1000}, for the transmission ringdown waveform $\Psi_{\text{F}}^{\text{G}}(u)$, the mismatch function increases with the increase of $\sigma$, regardless of $u_{\text{max}}$. These results demonstrate that the larger the width of the Gauss bump source, the more clearly it reflects the difference in the waveform between the piecewise step potential and the R-W potential. Here, the amplitude of the initial bump increases as $\sigma$ decreases [cf. Eq. (\ref{initial_Gauss_bump})], while the integration of the initial condition remains constant. If such integration can be regarded as energy, then our results will indicate that under given energy of the initial bump, the more dispersed the distribution of the bump [or in other words, the more low-frequency components there are in the initial bump, as can be seen from Eq. (\ref{T_G_omega})], the more sensitive it is to reflect tiny differences in the effective potential from the ringdown waveform aspect.


\section{Conclusions and discussion}\label{sec: conclusions}
In this work, we investigate the stability of waveforms by comparing the time-domain waveforms associated with the piecewise step approximation of the R-W potential to those of the original R-W potential. The difference between the R-W potential and its piecewise step approximation is interpreted as an external perturbation to the Schwarzschild background, whose strength can be controlled by the number of steps $N_{\text{st}}$. A larger $N_{\text{st}}$ yields a closer approximation to the original R-W potential. The construction of the piecewise step potential is described in Sec. \ref{Piecewise_step_approximation}. Notably, unlike the methods adopted in~\cite{Nollert:1996rf,Daghigh:2020jyk,Xie:2025jbr}, our approach employs Chebyshev-Lobatto grids to partition the effective potential range, $[0,V_{\text{R-W}}(x_m)]$. This method enables the piecewise step potential to approximate the R-W potential more efficiently.

In Sec. \ref{sec: QNMs}, we review the problem of QNMs for the piecewise step potential~\cite{Nollert:1996rf}. Following our previous work~\cite{Xie:2025jbr}, we employ the transfer matrix method to derive the QNM condition $A_{\mathrm{in}}(\omega_{n})=0$. In practice, we first locate the minima of $\ln|A_{\mathrm{in}}(\omega)|$ in the complex plane and then use these locations as initial guesses of the Muller method to compute the QNM spectra. This procedure is computationally efficient and numerically accurate, particularly suitable for the analytic expressions generated by the transfer matrix method for piecewise potentials. The final results are displayed in Fig. \ref{QNM_spectra}, and they are overall consistent with those reported in~\cite{Nollert:1996rf}: A clear bifurcation structure is observed, and the imaginary part $\mathrm{Im}(\omega_{n})$ decreases slowly as the overtone number grows, which is dramatically different from the Schwarzschild case. This indicates that the QNM spectra can undergo disproportionate shifts under small deformations of the effective potential, exhibiting significant instability.

It is generally understood that the ringdown signal can be expressed as a superposition of QNMs: $\sum_{n} E_{n} T_{n} \mathrm{e}^{-\mathrm{i} \omega_{n} t}$, indicating that the waveform is not solely characterized by the QNM spectra, as the excitation factor $E_{n}$ and the source factor $T_{n}$ also play a crucial role in shaping the overall time-domain signal. The excitation factor, like the QNM spectra, are only determined by the geometry of the background spacetime, and there exists a one-to-one correspondence between them. The detailed derivations are presented in Sec. \ref{sec: EFs}. The results $|E_{n}|$ of piecewise step approximated potential are shown in Fig. \ref{abs_EFs_piecewise}.

It has been verified in many works that the ringdown waveform in the time-domain is insensitive to small deformations of the effective potential. Since the initial condition also affects the time-domain waveform, it is natural to consider their influences when studying the stability of waveforms. In particular, whether the general conclusion of waveform stability breaks down for certain choices of initial condition, and what kinds of initial conditions can better detect the small deformations of the effective potential outside the Schwarzschild black hole. Here we employ the mismatch function (\ref{mismatch}) to quantitatively characterize the stability. These studies are investigated in Sec. \ref{sec: stability_of_waveforms}.

For simplicity, we consider the case of an observer at spatial infinity, in which the ringdown waveform is given by Eq. (\ref{general_solution_time_domain}). With this simplification, for some specific physical cases, where the initial Gauss bump locates in spatial infinity or near the event horizon, the Green's function method allows us to compute Eq. (\ref{general_solution_time_domain}) analytically, and we obtain Eq. (\ref{waveform_sourced_Gauss_infinity}) and Eq. (\ref{waveform_sourced_Gauss_event}). When $\sigma=0$, Gauss bump reduces to the Dirac delta function, and then one gets the ringdown waveforms (\ref{fundamental_ringdown_H}) and (\ref{fundamental_ringdown_F}). First, we study the behaviors of the transmission amplitude $F(\omega)$ and reflection amplitude $H(\omega)$. As our results shown, their magnitudes, i.e., $|H(\omega)|$ and $|F(\omega)|$, indeed converge to those of the R-W potential as $N_{\text{st}}$ increases [see Fig. \ref{Abs_H_omega} and Fig. \ref{Abs_F_omega}]. However, their corresponding phases exhibit different behavior. As shown in Fig. \ref{abs_H_omega_and_delta_omega}, the phase of the reflection amplitude, $\delta(\omega)$, develops significant instability in the high-frequency regime. Only in the low-frequency domain does $\delta_{P}(\omega)$ converge to $\delta_{S}(\omega)$ with increasing $N_{\text{st}}$, while a relative phase shift of $2\pi$ persist in the vicinity of zero-frequency. As presented in Fig. \ref{abs_F_omega_and_eta_omega}, the deviation between $\eta_{P}(\omega)$ and $\eta_{S}(\omega)$ remain small in the high-frequency regime but becomes substantial in the low-frequency regime. Similar to the behavior of $\delta(\omega)$, $\eta_{P}(\omega)$ also approaches $\eta_{S}(\omega)$ as $N_{\text{st}}$ increases, while in the vicinity of zero frequency they differ by $\pi$.

For the general Gauss bump source $(\sigma>0)$, we further investigate the stability of waveform affected by the bump width $\sigma$. As the piecewise step potential provides a closer approximation to the R-W potential, the agreement between the corresponding waveforms consistently improves, indicating that the time-domain waveform is indeed stable. Moreover, our results show that when considering long durations of the waveforms, waveforms initiated with broader bumps (or equivalently, those with more low-frequency components) exhibit a larger mismatch with the Schwarzschild case than those narrower initial Gauss bumps. This suggests that waveforms excited by broader initial bumps are more sensitive to small deviations in the effective potential outside the black hole. Hence, using broader initial bumps as probes of the black hole's exterior, or identifying astrophysical events whose initial data resembles such broader bumps may provide a more effective strategy for detecting subtle departure from the theoretical ideal potential.

So far, we have studied the waveform stability for the piecewise step approximation of R-W potential within the Gauss bump source. In a more physical situation, the source term $I(\omega,x)$ can also come from point particles orbiting around a black hole. This inspires us to consider the effect of non-smooth correction of the effective potential on the waveform from a point-like object source, which leaves for the future.

\section*{Acknowledgement}
This work is supported in part by the National Key Research and Development Program of China Grant No. 2020YFC2201501, in part by the National Natural Science Foundation of China under Grant No. 12475067 and No. 12235019. This work is also supported by the National Natural Science Foundation of China with grant No. 12505067.

\appendix
\section{Waveforms sourced by Gauss bump}\label{sec: waveforms_sourced_by_Gauss_bump}
In this appendix, we obtain a waveform which is sourced by a Gauss bump. In other words, we set up initial conditions such that
\begin{eqnarray}\label{initial_Gauss_bump}
    \Psi(t,x)\Big|_{t=0}=\frac{2A}{\sqrt{2\pi}\sigma}\exp\Big[-\frac{(x-x_s)^2}{2\sigma^2}\Big]\, ,\quad \frac{\partial\Psi}{\partial t}\Big|_{t=0}=0\, ,
\end{eqnarray}
with $A$ refers to the amplitude, $x_s$ refers to the position, and $\sigma$ refers to the width of the Gauss bump. It means that the source term $I(\omega,x)$ reads 
\begin{eqnarray}\label{I_Gauss}
    I(\omega,x)=\frac{2\mathrm{i}\omega A}{\sqrt{2\pi}\sigma}\exp\Big[-\frac{(x-x_s)^2}{2\sigma^2}\Big]\, .
\end{eqnarray}

For the first case, we assume that $x_s\gg x_m$, where $x_m$ is the extremum of the R-W potential. Since the support set of $I(\omega,x)$ is around $x_s$, where $x_s$ approaches the positive infinity, and the ``in'' solution on the support set is about $\tilde{\Psi}_{\text{in}}(\omega,x)=A_{\text{in}}(\omega)\mathrm{e}^{-\mathrm{i}\omega x}+A_{\text{out}}(\omega)\mathrm{e}^{\mathrm{i}\omega x}$. Accordingly, the source factor (\ref{tilde_Psi_G_and_tilde_Psi_T}) is approximated as
\begin{eqnarray}
    \tilde{\Psi}_{\text{T}}(\omega)=\frac{2\mathrm{i}\omega A}{\sqrt{2\pi}\sigma A_{\text{out}}(\omega)}\int_{-\infty}^{+\infty}\Big(A_{\text{in}}(\omega)\mathrm{e}^{-(x^{\prime}-x_s)^2/(2\sigma^2)}\mathrm{e}^{-\mathrm{i}\omega x^{\prime}}+A_{\text{out}}(\omega)\mathrm{e}^{-(x^{\prime}-x_s)^2/(2\sigma^2)}\mathrm{e}^{\mathrm{i}\omega x^{\prime}}\Big)\mathrm{d}x^{\prime}\, .
\end{eqnarray}
Therefore, when the condition that both $x$ and $x_s$ tend towards positive infinity establishs, Eq. (\ref{general_solution_time_domain}) becomes
\begin{eqnarray}\label{waveform_sourced_Gauss_infinity}
    \Psi(t,x)&=&\frac{1}{2\pi}\int_{-\infty}^{+\infty}\mathrm{d}\omega \mathrm{e}^{-\mathrm{i}\omega (t-x)}\frac{A}{\sqrt{2\pi}\sigma}\Bigg[\int_{-\infty}^{+\infty}\mathrm{e}^{-(x^{\prime}-x_s)^2/(2\sigma^2)}\mathrm{e}^{-\mathrm{i}\omega x^{\prime}}\mathrm{d}x^{\prime}\Bigg]\nonumber\\
    &&+\frac{1}{2\pi}\int_{-\infty}^{+\infty}\mathrm{d}\omega \mathrm{e}^{-\mathrm{i}\omega (t-x)}\frac{A A_{\text{out}}(\omega)}{\sqrt{2\pi}\sigma A_{\text{in}}(\omega)}\Bigg[\int_{-\infty}^{+\infty}\mathrm{e}^{-(x^{\prime}-x_s)^2/(2\sigma^2)}\mathrm{e}^{\mathrm{i}\omega x^{\prime}}\mathrm{d}x^{\prime}\Bigg]\nonumber\\
    &=&\frac{1}{2\pi}\int_{-\infty}^{+\infty}\mathrm{d}\omega \mathrm{e}^{-\mathrm{i}\omega (t-x)}A\exp\Big[-\frac{1}{2}\omega\Big(2\mathrm{i} x_s+\sigma^2\omega\Big)\Big]\nonumber\\
    &&+\frac{1}{2\pi}\int_{-\infty}^{+\infty}\mathrm{d}\omega \mathrm{e}^{-\mathrm{i}\omega (t-x)}AH(\omega)\exp\Big[\frac{1}{2}\omega\Big(2\mathrm{i} x_s-\sigma^2\omega\Big)\Big]\nonumber\\
    &=&\frac{A}{\sqrt{2\pi}\sigma}\mathrm{e}^{-(t-x+x_s)^2/(2\sigma^2)}+\frac{A}{2\pi}\int_{-\infty}^{+\infty}\mathrm{d}\omega \mathrm{e}^{-\mathrm{i}\omega (t-x-x_s)}H(\omega)T_{\text{G}}(\omega)\, ,
\end{eqnarray}
where
\begin{eqnarray}\label{T_G_omega}
    T_{\text{G}}(\omega)=\exp\Big(-\frac{\sigma^2\omega^2}{2}\Big)\, .
\end{eqnarray}
The physical meaning of the waveform (\ref{waveform_sourced_Gauss_infinity}) is quite obvious. The first part is the waveform directly received by an observer at infinity. The second part is the waveform received by an observer at infinity after being reflected by the effective potential, which is precisely the ringdown part. As the width $\sigma\to0$, assuming $A$ is independent of $\sigma$, we have $I=2\mathrm{i}\omega\delta(x-x_s)$ from Eq. (\ref{I_Gauss}). Therefore, the waveform in Eq. (\ref{waveform_sourced_Gauss_infinity}) will be
\begin{eqnarray}\label{waveform_sourced_delta_infinity}
    \Psi(t,x)=A\delta(t-x+x_s)+A\Psi_{\text{H}}(t-x_s,x)\, .
\end{eqnarray}
The first part of the above waveform (\ref{waveform_sourced_delta_infinity}) is a delta function. The second part is the waveform obtained from the reflection amplitude, which is studied in~\cite{Oshita:2024wgt,Oshita:2025ibu,Rosato:2025byu,Rosato:2024arw}. Note that comparing $A\Psi_{\text{H}}(t-x_s,x)$ and Eq. (\ref{fundamental_ringdown_H}), we know the former is delayed by $x_s$.

For the second case, we assume that $x_s\ll x_m$. Since the support set of $I(\omega,x)$ is around $x_s$, where $x_s$ approaches the negative infinity (event horizon), and the ``in'' solution on the support set is about $\tilde{\Psi}_{\text{in}}(\omega,x)=\mathrm{e}^{-\mathrm{i}\omega x}$. Accordingly, the source factor (\ref{tilde_Psi_G_and_tilde_Psi_T}) is approximated as
\begin{eqnarray}
    \tilde{\Psi}_{\text{T}}(\omega)=\frac{2\mathrm{i}\omega A}{\sqrt{2\pi}\sigma A_{\text{out}}(\omega)}\int_{-\infty}^{+\infty}\mathrm{e}^{-(x^{\prime}-x_s)^2/(2\sigma^2)}\mathrm{e}^{-\mathrm{i}\omega x^{\prime}}\mathrm{d}x^{\prime}\, .
\end{eqnarray}

Therefore, when the condition that $x$ tends towards positive infinity and $x_s$ tends towards negative infinity establishes, Eq. (\ref{general_solution_time_domain}) becomes
\begin{eqnarray}\label{waveform_sourced_Gauss_event}
    \Psi(t,x)&=&\frac{1}{2\pi}\int_{-\infty}^{+\infty}\mathrm{d}\omega \mathrm{e}^{-\mathrm{i}\omega (t-x)}\frac{A}{\sqrt{2\pi}\sigma A_{\text{in}}(\omega)}\Bigg[\int_{-\infty}^{+\infty}\mathrm{e}^{-(x^{\prime}-x_s)^2/(2\sigma^2)}\mathrm{e}^{-\mathrm{i}\omega x^{\prime}}\mathrm{d}x^{\prime}\Bigg]\nonumber\\
    &=&\frac{1}{2\pi}\int_{-\infty}^{+\infty}\mathrm{d}\omega \mathrm{e}^{-\mathrm{i}\omega (t-x)}\frac{A}{A_{\text{in}}(\omega)}\exp\Big[-\frac{1}{2}\omega\Big(2\mathrm{i} x_s+\sigma^2\omega\Big)\Big]\nonumber\\
    &=&\frac{A}{2\pi}\int_{-\infty}^{+\infty}\mathrm{d}\omega \mathrm{e}^{-\mathrm{i}\omega (t-x+x_s)}F(\omega)T_{\text{G}}(\omega)\, .
\end{eqnarray}
The physical meaning of Eq. (\ref{waveform_sourced_Gauss_event}) is that a waveform is received by an observer at infinity after being transmitted by an effective potential. As the width $\sigma\to0$, assuming $A$ is independent of $\sigma$, we have $I=2\mathrm{i}\omega\delta(x-x_s)$ from Eq. (\ref{I_Gauss}). Therefore, the waveform in Eq. (\ref{waveform_sourced_Gauss_event}) will be
\begin{eqnarray}\label{waveform_sourced_delta_event}
    \Psi(t,x)=A\Psi_{\text{F}}(t+x_s,x)\, .
\end{eqnarray}
Note that comparing $A\Psi_{\text{F}}(t+x_s,x)$ and Eq. (\ref{fundamental_ringdown_F}), we know the former is advanced by $x_s$.


\bibliography{reference}
\bibliographystyle{apsrev4-1}

\end{document}